\catcode`\@=11
\newif\if@fewtab\@fewtabtrue

{\count255=\time\divide\count255 by 60
\xdef\hourmin{\number\count255}
\multiply\count255 by-60\advance\count255 by\time
\xdef\hourmin{\hourmin:\ifnum\count255<10 0\fi\the\count255}}
\def\ps@draft{\let\@mkboth\@gobbletwo
    \def\@oddhead{}
    \def\@oddfoot
       {\hbox to 7 cm{$\scriptstyle Draft\ version:\ \draftdate$
       \hfil}\hskip -7cm\hfil\rm\thepage \hfil}
    \def\@evenhead{}\let\@evenfoot\@oddfoot}

\def\ceqno{\global\@fewtabfalse
    \ifcase\@eqcnt \def\@tempa{& & &}\or \def\@tempa{& &}
      \or \def\@tempa{&}
      \or\def\@tempa{}\fi\@tempa
{\rm(\theequation)}}

\def\aeqno#1{\global\@fewtabfalse
    \ifcase\@eqcnt \def\@tempa{& & &}\or \def\@tempa{& &}
      \or \def\@tempa{&}
      \or\def\@tempa{}\fi\@tempa
{\rm(\theequation,#1)}}

\def\label#1{\ifnum\draftcontrol=1
 \global\def\draftnote{$\scriptstyle #1$}\fi
 \@bsphack\if@filesw {\let\thepage\relax
   \def\protect{\noexpand\noexpand\noexpand}%
\xdef\@gtempa{\write\@auxout{\string
      \newlabel{#1}{{\@currentlabel}{\thepage}}}}}\@gtempa
   \if@nobreak \ifvmode\nobreak\fi\fi\fi
  \@esphack}

\def\alabel#1#2{\label{#1}\global\@fewtabfalse
    \ifcase\@eqcnt \def\@tempa{& & &}\or \def\@tempa{& &}
      \or \def\@tempa{&}
      \or\def\@tempa{}\fi\@tempa
{\hbox to 3cm{\phantom{\rm(\theequation,#2)}
\draftnote \hfil}\hskip -3cm {\rm(\theequation,#2)}}}

\def\clabel#1{\label{#1}\global\@fewtabfalse
    \ifcase\@eqcnt \def\@tempa{& & &}\or \def\@tempa{& &}
      \or \def\@tempa{&}
      \or\def\@tempa{}\fi\@tempa
{\hbox to 3cm{\phantom{\rm(\theequation)}
\draftnote \hfil}\hskip -3cm{\rm(\theequation)}}}

\def\eqnarray{\def\draftnote{{}}\global\@fewtabtrue
\stepcounter{equation}\let\@currentlabel=\theequation
\global\@eqnswtrue
\global\@eqcnt\z@\tabskip\@centering\let\\=\@eqncr
$$\halign to \displaywidth\bgroup\@eqnsel\hskip\@centering\@eqcnt\z@
  $\displaystyle\tabskip\z@{##}$&\global\@eqcnt\@ne
  \hskip 1\arraycolsep \hfil${##}$\hfil
  &\global\@eqcnt\tw@ \hskip 1\arraycolsep
$\displaystyle\tabskip\z@{##}$
\hfil  \tabskip\@centering&\global\@eqcnt\thr@@\llap{##}\tabskip\z@
\cr}

\def\endeqnarray{\@@eqncr\egroup
      \global\advance\c@equation\m@ne$$\global\@ignoretrue}

\def\@eqnnum{\hbox to 3cm{\phantom{\rm(\theequation)} \draftnote
                         \hfil}\hskip -3cm {\rm(\theequation)}}

\def\@@eqncr{\let\@tempa\relax
    \ifcase\@eqcnt \def\@tempa{& & &}\or \def\@tempa{& &}
      \or \def\@tempa{&}
      \or\def\@tempa{}
\fi\@tempa
\if@eqnsw
\if@fewtab\@eqnnum\fi
\stepcounter{equation}\fi\global
\@eqnswtrue\global\@eqcnt\z@\global\@fewtabtrue\cr}

\def\draftcite#1{\ifnum\draftcontrol=1#1\else{}\fi}

\def\@lbibitem[#1]#2{\item{}\hskip -3cm \hbox to 2cm
{\hfil$\scriptstyle\draftcite{#2}$}\hskip
1cm[\@biblabel{#1}]\if@filesw
     {\def\protect##1{\string ##1\space}\immediate
      \write\@auxout{\string\bibcite{#2}{#1}}}\fi\ignorespaces}

\def\@bibitem#1{\item\hskip -3cm \hbox to 2cm
{\hfil $\scriptstyle\draftcite{#1}$}\hskip 1cm
\if@filesw \immediate\write\@auxout
       {\string\bibcite{#1}{\the\value{\@listctr}}}\fi\ignorespaces}

 \def\nsection#1{\vskip 0.1cm\no{\large\bf LECTURE
         #1}\def\thesection{#1} \setcounter{equation}{0}}

 \def\nconclusions#1{\vskip 1cm\no{\large\bf CONCLUSIONS
         #1}\def\thesection{#1} \setcounter{equation}{0}}

 \def\nreferences#1{\vskip 1.5cm\no{\large\bf REFERENCES
         #1}\def\thesection{#1} \setcounter{equation}{0}}

\font\tendl=msbm10  scaled \magstep1
\font\sevendl=msbm7 scaled \magstep1
\font\fivedl=msbm5 scaled \magstep1
\font\tengl=eufm10  scaled \magstep1
\font\sevengl=eufm7 scaled \magstep1
\font\fivegl=eufm5 scaled \magstep1

\newfam\dlfam  
\textfont\dlfam=\tendl \scriptfont\dlfam=\sevendl
\scriptscriptfont\dlfam=\fivedl
\newfam\glfam  
\textfont\glfam=\tengl \scriptfont\glfam=\sevengl
\scriptscriptfont\glfam=\fivegl

\def\draftdate{\number\month/\number\day/\number\year\ \ \ \hourmin }

\global\def\draftcontrol{0}
\catcode`\@=12
\def\tilde{\widetilde}
\def\hat{\widehat}

\documentstyle[12pt,epsf]{article}

\def\theequation{{\thesection.\arabic{equation}}}
\newcommand{\be}{\begin{eqnarray}}
\newcommand{\en}{\end{eqnarray}\vs 0.5 cm}
\newcommand{\non}{\nonumber}
\newcommand{\no}{\noindent}
\newcommand{\vs}{\vskip}

\newcommand{\un}{\underline}

\newcommand{\Nx}{{\bf x}}

\newcommand{\Ny}{{\bf y}}

\newcommand{\Nv}{{\bf v}}
\newcommand{\Nw}{{\bf w}}
\newcommand{\Nk}{{\bf k}}
\newcommand{\Nna}{{\bf \nabla}}
\newcommand{\NR}{{{\bf R}}}
\newcommand{\NV}{{{\bf V}}}
\newcommand{\bm}[1]{\mbox{\boldmath $#1$}}
\newcommand{\qq}{\begin{eqnarray}}

\newcommand{\da}{\partial}
\newcommand{\ov}{\overline}
\newcommand{\ee}{{\rm e}}

\newcommand{\qqq}{\end{eqnarray}}

\newcommand{\CK}{{\cal K}}

\newcommand{\CM}{{\cal M}}

\newcommand{\CO}{{\cal O}}
\newcommand{\CP}{{\cal P}}

\newcommand{\CS}{{\cal S}}
\newcommand{\CT}{{\cal T}}

\newcommand{\s}{\hspace{0.05cm}}
\newcommand{\m}{\hspace{0.025cm}}

\newcommand{\hf}{{_1\over^2}}

\def\refce{\small\medskip\smallskip\noindent\hangindent\parindent}

\begin{document}

\title{{\bf Soluble models \,of \,turbulent advection}\footnote{Lectures
given at the workshop ``Random Media 2000'', M\c{a}dralin by Warsaw,
June 19-26, 2000}}

\author{\ \\Krzysztof Gaw\c{e}dzki\\ \ \\
C.N.R.S., Laboratoire de Physique, ENS-Lyon,\\46, All\'ee d'Italie,
F-69364 Lyon, France}

\date{ }

\maketitle

\addtocounter{section}{0}


\noindent The understanding of developed turbulence, a long
standing challenge for mathematical physics, has entered into the
third millennium as an unsolved problem. It poses basic questions
concerning both the behavior of solutions of hydrodynamical
equations and the basic principles of statistical mechanics of
systems out of equilibrium. Although we seem far away from the
definite answers to these million (or more) dollar questions
(Fefferman, 2000), some insight may be gained by studying simpler
models showing similar behaviors, often themselves not devoided of
direct physical interest. One of such problems concerns the
passive transport of scalar quantities such as temperature or
tracer $\,\theta(t,\Nx)\,$ and pollutant or dye density
$\,\rho(t,\Nx)\,$ by random flows. The flows are described by a
simple random ensemble of velocities $\,\Nv(t,\Nx)\,$ that is
considered given. Such approach ignores the back-reaction of the
scalar on the velocity dynamics as well as many details of that
dynamics. It incorporates, however, the basic phenomenological
property of velocities exhibiting developed turbulence: their
(approximate) statistical scaling \qq \Nv(t,\Nx)-\Nv(t,\Ny)\ \
\sim\ \ |\Nx-\Ny|^\alpha \label{stsc} \qqq signaled by the scaling
behavior of the expectations of the powers of the velocity
differences, the so called velocity {\bf structure functions}, \qq
{\bf E}\ |\Nv(t,\Nx)-\Nv(t,\Ny)|^N\ \ \simeq\ \ |\Nx-\Ny|^{
\zeta_{_N}}\qqq with $\,\zeta_{_N}\,$ (for low $N$) not far from
the mean field theory (Kolmogorov, 1941) value $\,\zeta^{\rm
Kol}_{_N}=N/3\,$ that would correspond to $\,\alpha=1/3$. \,The
time-evolution of the scalar is described by the {\bf
advection-diffusion equation} \qq
\da_t\theta\s+\s(\Nv\cdot\Nna)\theta\s- \s\kappa\Nna^2\theta\s=\s
f\,, \label{ps} \qqq where $\,\kappa\,$ is the (molecular)
diffusivity constant and $\,f(t,\Nx)\,$ denotes a scalar source
that one may also take random. Given the distributions of the
velocities $\,\Nv\,$ and of the sources $\,f$, \,one inquires
about the statistics of solutions $\,\theta\,$ of the above
equation. As we shall see, very simple distributions of $\,\Nv\,$
and $\,f\,$ lead to stationary statistical states of scalar
exhibiting many features observed not only in realistic turbulent
transport but also in statistics of turbulent velocities, in the
atmosphere, in aerodynamical tunnels or in sea channels. Those
features include persistent dissipation of energy, energy cascades
from large scales to small ones or vice versa and intermittency.
Simple models allow a better understanding of the origin of such
behaviors and permit to draw some general conclusions, see
(Falkovich, Gawedzki \& Vergassola, 2001) for an extensive review
and bibliography. \vskip 1.5cm

\nsection{1}
\vskip 0.3cm

\noindent ({\small transport of scalars by hydrodynamical flow; the
role of fluid particle dynamics, single particle diffusion versus
two-particle dispersion})
\vskip 0.8cm

\noindent 1.1. \ {\bf Solution of the advection-diffusion equation, fluid
particles}
\vskip 0.5cm

\noindent It is easy to solve the linear equation (\ref{ps}) describing
the scalar evolution in $\,d\,$ space dimensions when the velocity field
is sufficiently smooth.
\vskip 0.3cm

\noindent (i).\,\, \ For $\kappa=0$ and $f=0$, the scalar $\,\theta(t,\Nx)\,$
is simply constant along the characteristics
\qq
{d{\bm x}\over ds}\ =\ \Nv(s,{\bm x})
\label{ode}
\qqq
which describe the (Lagrangian) trajectories of fluid
particles. In other words, the scalar is carried by the flow:
\qq
\theta(t,\Nx)\ =\ \theta(s,{\bm x}_{t,\Nx}(s))\,,
\qqq
where $\,{\bm x}_{t,\Nx}(s)\,$ is the Lagrangian trajectory
that passes at time $\,t\,$ through point $\,\Nx$.
Note that the forward evolution of $\,\theta\,$ corresponds
to the backward Lagrangian flow.
\vskip 0.3cm

\noindent (ii). \,\,\ In the presence of the source $\,f$,
\,the scalar is also created or depleted along the trajectory:
\qq
\theta(t,\Nx)\ =\,\theta(s,{\bm x}_{t,\Nx}(s))\,+\,\int\limits_s^t
f(\sigma,{\bm x}_{t,\Nx}(\sigma))\,\m d\sigma\,.
\label{sol}
\qqq
\vskip 0.3cm

\noindent (iii). Finally, when $\kappa\not=0$, $\,{\bm
x}_{t,\Nx}(s)$ should be taken as the solution of the stochastic
ODE \qq d{\bm x}\ =\ \Nv(s,{\bm x})\,ds\,+\,\sqrt{2\kappa}\,\,d\Nw
\label{sode} \qqq for the Lagrangian trajectories perturbed by the
$d$-dimensional Brownian motion $\,\Nw(s)\,$ and the right hand
side of eq.\,\,(\ref{sol}) should be averaged over $\Nw$: \qq
\theta(t,\Nx)\ =\ {\bf E}_{_{\Nw}}\,\Big(\,\theta(s,{\bm
x}_{t,\Nx}(s))\, +\,\int\limits_s^tf(\sigma,{\bm
x}_{t,\Nx}(\sigma))\,\m d\sigma\,\Big). \label{ssol} \qqq \vskip
0.3cm

\noindent It is then clear that the statistics of
$\,\theta(t,\Nx)\,$ is determined by the statistics of the (noisy)
Lagrangian trajectories. The latter may be captured in two steps.
First, we may consider the transition probabilities for the Markov
process given by the solution of the stochastic equation
(\ref{sode}) in a fixed velocity field: \qq P(\Nv|t,\Nx;s,d\Ny)\
=\ {\bf E}_{_\Nw}\ \delta(\Ny-{\bm x}_{t,\Nx}(s)) \,d\Ny\,. \qqq
Note that the solution (\ref{ssol}) may be rewritten as \qq
\theta(t,\Nx)\ = \ \int P(\Nv|t,\Nx;s,d\Ny)\,\,\theta(s,\Ny)
\,+\,\int\limits_s^td\sigma\int
P(\Nv|t,\Nx;\sigma,d\Ny)\,\,f(\sigma,\Ny)\label{han0} \qqq or, in
the more handy operator notation, as \qq \theta(t)\ =\
P(\Nv|t,s)\,\,\theta(s)\ + \ \int_s^tP(\Nv|t,\sigma)
\,\,f(\sigma)\,\,d\sigma\,. \label{han} \qqq The latter formuli
make sense for $\,\kappa>0\,$ also in rough (non-Lipschitz)
velocities when eq.\,\,(\ref{ode}) does not have unique solutions.
Second, in order to take account of the velocity fluctuations, we
may consider the joint probability distributions of $\,N\,$
Lagrangian particles \qq
\CP_{_N}(\un{t},\un{\Nx};\un{s},d\un{\Ny})\ =\ {\bf E}\
\prod\limits_{n=1}^N P(\Nv|t_n,\Nx_n;s_n,d\Ny_n)\,, \label{san}
\qqq where $\,\un{t}=(t_{_1}\dots,t_{_N})$,\
$\,\un{\Nx}=(\Nx_{_1},\dots, \Nx_{_N})$, \,etc. and $\,{\bf E}$
stands for the average over the velocity ensemble. We shall be
interested in the random ensembles of velocities that are
stationary, homogeneous and isotropic, i.e.\,\,such that the time
and space translations and rotations
$\,\Nv(t,\Nx)\mapsto\Nv(t+t_0,R_0\Nx+\Nx_0)\,$ are implemented by
the measure preserving action of the corresponding groups on the
probability space of velocities. \vskip 0.7cm

\noindent 1.2. \ {\bf Single-particle diffusion and Richardson dispersion
of two particles}
\vskip 0.5cm

\noindent What is the statistical behavior of fluid particles in
stationary, homogeneous and isotropic turbulent velocities? The
rough answers are as follows. For a single particle, one expects a
diffusive behavior for sufficiently long times. Note that \qq {\bm
x}_{0,\Nx}(t)-\Nx\,=\,\int\limits_0^t\Nv(\sigma,{\bm
x}_{0,\Nx}(\sigma))
\,d\sigma\,\equiv\,\int\limits_0^t\Nv_{_L}(\sigma,\Nx)\,d\sigma
\label{int} \qqq for solutions of eq.\,\,(\ref{ode}), where
$\,\Nv_{_L}(\sigma,\Nx)\,$ denotes the the velocity along the
Lagrangian trajectory passing at time zero through $\,\Nx$, \,the
so called Lagrangian velocity. For fixed $\,\Nx$,
$\,\Nv_{_L}(\sigma,\Nx)\,$ is a stationary process as long as
velocities are incompressible. It has zero expectation if the
similar property holds for the (Eulerian) velocity
$\,\Nv(\sigma,\Nx)\,$ with which it coincides at time zero. If
$\,\Nv_{_L}(s,\Nx)\,$ has temporal correlations that decay fast
enough, then the integral in (\ref{int}) falls under the Central
Limit Theorem behaving effectively as a sum of many independent
equally distributed random variables. As a result,
$\,\mu^{-\hf}{\bm x}_{0,\Nx}(\mu t)\,$ tends when
$\,\mu\to\infty\,$ to a Brownian motion and \qq \CP_1(0,\Nx;\mu
t,d(\mu^\hf\Ny))\ \,\ \mathop{\longrightarrow}
\limits_{\mu\to\infty}\ \,\ \ee^{\,\hf
t\,D_0\,\nabla^2}\hspace{-0.08cm}
(\Nx,\Ny)\,d\Ny\,=\,{_1\over^{(4\pi\,D_0 t)^{d/2}}}\,
\ee^{-(\Nx-\Ny)^2/(4\,D_0t)}\,d\Ny\,, \label{1psl} \qqq i.e.\,\,it
becomes the transition probability of diffusion. The diffusion
constant is given by the formula (Taylor, 1921) \qq
D_0\,=\,\int\limits_0^\infty{\bf
E}\,\,\Nv_{_L}(0,\Nx)\cdot\Nv_{_L}(\sigma,\Nx) \,\,d\sigma\,.
\label{tayl} \qqq Still, the decay of temporal correlations of the
Lagrangian velocity is a nontrivial fact, see (Fannjiang \&
Papanicolaou, 1996) or (Majda \& Kramer, 1999). It does not
automatically follow from similar property of the (Eulerian)
velocity $\,\Nv(\sigma,\Nx)\,$ with fixed $\Nx$ and may fail
altogether in specially prepared velocity ensembles. \vskip 0.4cm
 For two particles, the important quantity to study is the 2-particle
separation $\,{\bm\rho}(t)\,$ defined as the difference $\,{\bm
x}_{0,\Nx_2}(t)-{\bm x}_{0,\Nx_1}(t)\,$ of particle positions. It
satisfies the equation \qq {d{\bm\rho}\over ds}\ =\ \Nv(s,{\bm
x}_{0,\Nx_1}+\bm\rho(s))- \Nv(s,{\bm x}_{0,\Nx_1}(s))\,
\label{sep} \qqq with the initial condition
$\,\bm\rho(0)=\Nx_2-\Nx_1$. \,To get a rough idea about the
behavior of the particle separation, let us first consider small
separations in smooth slowly varying velocities where the right
hand side may be approximated by linear expression
$\,A\,{\bm\rho}(s)$. This leads to a solution \qq
{\bm\rho}(t)\,\simeq\,\ee^{A\,t}\,{\bm\rho}(0) \qqq with the
exponential growth if $\,A\,$ has eigenvalues with positive real
part (positive {\bf Lyapunov exponents}) that signal the sensitive
dependence on initial conditions usually considered as a
definition of {\bf chaos}. Notice that although the nearby
trajectories separate exponentially in this case, the very close
trajectories take long time to separate and infinitesimally close
ones never separate. As the result, the trajectories are still
labeled in a continuous way by their initial positions. Such fluid
particle trajectory behavior pertains to the so called Batchelor
regime of turbulent flows corresponding to short scales dominated
by viscous effects, \vskip 0.3cm

Suppose now that we solve the equation (\ref{sep}) in the regime
where the velocity difference behaves like $\,\rho(s)^\alpha\,$
with $\,\alpha<1$. \,Such scaling behaviors are observed in the
inertial range of scales of turbulent flows where viscous and
stearring effects are negligible, Passing to the scalar version of
eq.\,\,(\ref{sep}), we obtain \qq {d{\rho}^2\over ds}\ =\
2{d{\bm\rho}\over ds}\cdot{\bm\rho} \ \propto\ \rho^{\alpha+1}
\label{alp0}\qqq and, ignoring again the time and point dependence
as well as the statistical fluctuations in the proportionality
constants, we obtain \qq \rho(t)|^{1-\alpha}\ \simeq\
\rho(0)^{1-\alpha}\,+ \,{\rm const}.\,t\,. \label{alp} \qqq This
very rough estimate allows us to expect a power law growth of the
2-particle {\bf dispersion} (the distance between two particles),
wiping out the memory of the initial separation. In particular,
this would mean that infinitesimally close trajectories would
still separate in a finite time, unlike in smooth velocities,
leading to a spontaneous randomness in the Lagrangian flow at
$\,\kappa=0$. \,Of course, in non-smooth velocities (e.g. in the
H\"{o}lder continuous ones) one should not expect existence of the
deterministic Lagrangian flow with trajectories labeled by initial
conditions since the assumptions of the theorem about the
uniqueness of solutions of the ODE (\ref{ode}) require Lipschitz
continuity of velocities in space. That type of situation pertains
to the behavior of turbulent flows at very high (ideally,
infinite) Reynolds numbers when the scaling behavior (\ref{stsc})
extends down to very small (infinitesimal) separations crossing
over to the Lipschitz behavior with $\,\alpha=1\,$ only at scales
where the viscous effects become important (of order of fractions
of millimeter in the turbulent atmosphere). The super-diffusive
behavior $\,\rho^2(t)\propto t^3\,$ corresponding to (\ref{alp})
with the Kolmogorov value $\,\alpha=1/3\,$ has, indeed, been
observed phenomenologically for the mean squared 2-particle
dispersion. It constitutes the content of the first quantitative
law of developed turbulence formulated in (Richardson, 1926) on
the bases of experimental data about the separation of
meteorological balloons and smoke particles. \vskip 0.3cm

The above arguments ignored the temporal dependence of velocity fields
which plays an important role. Nevertheless, the basic conclusion
about the possibility of finite time separation of arbitrary close
Lagrangian trajectories in spatially rough velocity fields holds
even if velocities are completely decorrelated (white) in time, as
we shall see below.
\vskip 0.3cm

The statistics of the separation of two fluid particles may be
captured by the relative transition amplitudes \qq {\cal P}^{\rm
rel}_2(t,\bm\rho_0;s,d\bm\rho)\ =\ \int{\cal P}_2(t,t,
\Nx_1,\Nx_1+\bm\rho_0);s,s,\Ny_1,d(\Ny_1+\bm\rho)) \label{rel}
\qqq where the integral is over the final position $\Ny_1$ of the
first particle. A strong version of the Richardson-type
super-diffusive behavior for large times may be formulated as the
statement about the existence of the limit of the rescaled process
$\,\mu^{-{1\over1-\alpha}}{\bm\rho} (\mu t)\,$ or of the limit \qq
\lim\limits_{\mu\to\infty}\ \CP^{\rm rel}_2(0,\bm\rho_0;\mu t,
d(\mu^{1\over1-\alpha}\bm\rho))\,. \label{Risl} \qqq Note the
difference with the expected diffusive scaling (\ref{1psl}) for a
single particle. Again, this type of scaling is not automatically
guaranteed by the statistical scaling of the Eulerian velocity
differences since there are many points where the naive mean-field
type arguments may go wrong. The super-diffusive behavior of the
2-particle dispersion is, in general, even harder to establish
than the diffusive behavior of a single particle. \vskip 1cm

\nsection{2}
\vskip 0.3cm

\noindent ({\small Kraichnan ensemble of velocities; Le Jan-Raimond
construction of Lagrangian particle processes; multi-particle statistics})
\vskip 0.8cm

\noindent It is important to have at the disposal a simple model
where the ideas about the behavior of fluid particles discussed
in the first lecture could be tested rigorously.
\vskip 0.7cm

\noindent 2.1. \ {\bf Kraichnan ensemble of velocities}
\vskip 0.5cm

\noindent Such a model has arisen from the work initiated in
(Kraichnan, 1968). Kraichnan proposed to consider a Gaussian
ensemble of velocities decorrelated in time but with the scaling
properties in space built in. Gaussian ensembles are completely
determined by the 1-point and 2-point functions. One assumes that
the 1-point function of $\,\Nv\,$ vanishes and that \qq {\bf E}\
\Nv^i(t,\Nx)\,\Nv^j(s,\Ny)\ =\ \delta(t-s)\,D^{ij}(\Nx-\Ny)\,,
\label{2pfv} \qqq where \qq D^{ij}(\Nx)\ =\
\int\limits_{|\Nk|<\eta^{-1}} (\delta^{ij}-{k^ik^j\over k^2})\,\,
{\ee^{\,i\,\Nk\cdot\Nx}\over(k^2+L^{-2})^{(\xi+d)/2}}\,\,d\Nk\,.
\label{dd0}\qqq Decomposing \qq
D^{ij}\,=\,D_0\,\delta^{ij}\,-\,d^{ij}(\Nx) \label{dd} \qqq with
the first term equal to $\,D^{ij}(0)$, \,it is not difficult to
see that
$\,d^{ij}(\Nx)\,\mathop{\longrightarrow}\limits_{|\Nx|\to\infty}\,
D_0\,\delta^{ij}\,$ with $\,D_0=\CO(L^\xi)$. \,At short distances
\qq d^{ij}(\Nx)\ =\
D_1\,[(d+1)\,\delta^{ij}\,|\Nx|^2\,-\,2\,x^ix^j]\ +\
\CO(|\Nx|^4)\qquad{\rm for}\ \ |\Nx|\ll\eta\,, \label{smr} \qqq
whereas \qq d^{ij}(\Nx)\ \
\mathop{\longrightarrow}\limits_{\eta\to0\atop L\to\infty}\ \
D_2\,[(d-1+\xi)\,\delta^{ij}\,|\Nx|^\xi\,-\xi\,x^ix^j\,
|\Nx|^{\xi-2}]\,. \label{lsf} \qqq The limiting scaling formula
(\ref{lsf}) approximates well $\,d^{ij}(\Nx)\,$ in the ``inertial
interval'' $\,\eta\ll\Nx\ll L$. The scale $\,\eta\,$ plays the
role of the ``viscous scale'' within which the fractional scaling
of $\,d^{ij}(\Nx)\,$ is replaced by a quadratic behavior and scale
$\,L\,$ of the ``integral scale'' on which velocities decorrelate.
The matrix $\,d(\Nx)\,$ describes the correlations of the velocity
differences. E.g. \qq {\bf E}\ (v^i(t,\Nx)-v^i(t,{\bf
0}))\,(v^j(s,\Nx)-v^j(s,{\bf 0}))\ =\ 2\,\delta(t-s)\
d^{ij}(\Nx)\,. \qqq The Kraichnan ensemble of velocities
incorporates on the statistical level the scaling properties
(\ref{stsc}) of velocities in the inertial interval
$\,\eta\ll|\Nx-\Ny|\ll L$. For $\,\eta>0$, \,the Gaussian measure
of the ensemble is supported by smooth velocities. This is not the
case, however, in the limiting case $\,\eta=0\,$ where the viscous
scale is set to zero, mimicking the inviscid or infinite Reynolds
number ensemble of turbulent velocities. In this case the ensemble
measure is supported on velocities that in their spatial behavior
are H\"{o}lder continuous with any exponent smaller than
$\,\xi/2\,$ whereas the velocities with the H\"{o}lder exponent
bigger than $\,\xi/2\,$ have measure zero. Of course, in their
temporal behavior, the Kraichnan velocities behave as white noise
or a derivative of the Brownian motion and thus are
distributional. We shall have then to modify our mean-field
arguments adapting it to such a case. \vskip 0.7cm

\noindent 2.2. \ {\bf Advection-diffusion equation in the Kraichnan
velocities}
\vskip 0.5cm

\noindent Let us consider first the finite-dimensional analog of
the advection-diffusion equation (\ref{ps}), \qq \dot{\theta}\ =\
\beta(t)\,\theta\,-a\,\theta\,, \label{dot} \qqq where
$\,\theta(t)\,$ takes values in $\,\NR^D\,$ and $\,\beta(t)\,$ is
a skew-symmetric matrix (a counterpart of the operator
$\,-\Nv\cdot \Nna$) \,and $\,a\,$ a positive symmetric matrix (a
counterpart of $\,-\kappa\Nna^2$). \,Suppose first that
$\,\beta\,$ is a smooth function of time. The solution of the
above linear equation has the form
$\,\,\theta(t)=P(\beta|t,s)\,\theta(s)\,\,$ with the propagator
$\,P(\beta|t,s)\,$ given by the time-ordered exponential \qq
P(\beta|t,s)&=&\CT\,\,\ee^{\int\limits_s^t(\beta(\sigma)-a)\,d\sigma}
\cr&=&\,\sum\limits_{n=0}^\infty\,\,\int\limits_{s\leq\sigma_1
\leq\sigma_2\leq\dots\leq \sigma_n\leq t}
\ee^{-(t-\sigma_n)\,a}\,\beta(\sigma_n)\,\,d\sigma_n
\,\,\ee^{-(\sigma_n-\sigma_{n-1})\,a}\,\dots\cr\cr
&&\hspace{3.3cm}\dots\,\beta(\sigma_2)\,d\sigma_2\,\,
\ee^{-(\sigma_2-\sigma_1)\,a}\,\beta(\sigma_1)\,\,d\sigma_1
\,\,\ee^{-(\sigma_1-s)\,a} \label{toi} \qqq for $\,t>s$. \,Another
way to express the same solution is by a limiting procedure: \qq
P(\beta|t,s)\ =\ \lim\limits_{{\rm
min}(\sigma_m-\sigma_{m-1})\searrow\,0} \
\ee^{\,\int\limits_{\sigma_n}^t(\beta(\sigma)-a)\,d\sigma}\,
\ee^{\,\int\limits_{\sigma_{n-1}}^{\sigma_n}(\beta(\sigma)-a)\,d\sigma}
\dots\,\ \ee^{\,\int\limits_{s}^{\sigma_1}(\beta(\sigma)-a)
\,d\sigma}\,. \label{top} \qqq Suppose now that $\,\beta(t)\,$ is
a white-noise Gaussian process with values in antisymmetric
matrices with mean zero and with the 2-point function \qq {\bf E}\
\beta^{\alpha\beta}(t)\,\beta^{\gamma\delta}(s)\ =\
\delta(t-s)\,C^{\alpha\beta,\gamma\delta}\,. \label{gfd} \qqq Now
$\,\beta\,d\sigma=dW\,$ is a differential of a Brownian
marix-valued motion $\,W(t)\,$ and the integrals in (\ref{toi})
become stochastic ones. Consequently, we fall into the standard
ambiguity with the choice of the convention for the latter. On the
other hand, the formula (\ref{top}) still makes sense with the
convergence taking place in any $\,L^p\,$ of the Gaussian process
with $\,p<\infty$, \,as it is not difficult to show. We shall take
it as the solution of the equation (\ref{dot}) for the white
$\,\beta(t)$. \,In terms of the standard conventions, this
corresponds to the Stratonovich prescription in (\ref{toi}) and
provides the solution of the Stratonovich stochastic ODE \qq
d\theta\ =\ dW\circ\theta\,-\,a\,\theta\,dt \qqq or, equivalently,
of the It\^{o} one: \qq d\theta\ =\
(dW)\,\theta\,-\,(a+c)\,\theta\,dt \qqq with
$\,c^{\alpha\beta}=-\hf\,C^{\alpha\gamma,\gamma\beta}\,$
(summation over $\gamma$!). The solution of the latter equation,
in turn, is given by the version of eq. (\ref{toi}) with $\,a\,$
replaced by $\,(a+c)\,$ and the integrals interpreted as the
It\^{o} stochastic ones and $\,\beta(\sigma)\,d\sigma\,$ as
$\,dW(\sigma)$. \vskip0.3cm

We shall then try to define the the solution of the
advection-diffusion equation (\ref{ps}) for white in time
velocities $\,v(t,\Nx)\,dt= d\NV(t,\Nx)\,$ as given by (\ref{han})
with $\,P(\Nv|t,s)\,$ represented by \qq
P(\Nv|t,s)&=&\CT\,\,\ee^{\,\int\limits_s^t(-d\NV(\sigma)\cdot\Nna+
\tilde\kappa\Nna^2\,d\sigma)} \cr\cr
&=&\sum\limits_{n=0}^\infty\,(-1)^n\,\int\limits_{s\leq\sigma_1
\leq\sigma_2\leq\dots\leq\sigma_n\leq t}
\ee^{\,(t-\sigma_n)\,\tilde\kappa\Nna^2}\,d\NV(\sigma_n)\cdot\Nna
\,\ee^{\,(\sigma_n-\sigma_{n-1})\,\tilde\kappa\Nna^2}\,\cr\cr
&&\dots\,d\NV(\sigma_2)\cdot\Nna\,
\ee^{\,(\sigma_2-\sigma_1)\,\tilde\kappa\Nna^2}\,
d\NV(\sigma_1)\cdot\Nna\,\ee^{\,(\sigma_1-s)\,\tilde\kappa\Nna^2}\,\equiv
\,\,\sum\limits_{n=0}^\infty P_n(\Nv|t,s)\,, \label{ito} \qqq with
the Ito stochastic integrals and $\,\tilde\kappa=\kappa+\hf D_0$.
The operator-valued white noise $\,-d\NV(t)\cdot\Nna$ plays the
role of $\,dW(t)$, the operator $\,\kappa\Nna^2\,$ the one of
$\,-a$, \,and $\,\hf D_0\nabla^2\,$ that of $\,-c$, \,as is easy
to figure out by comparing (\ref{2pfv}) and (\ref{dd}) with
(\ref{gfd}). \vskip 0.3cm

It is not difficult to check that each term of the series is well
defined in the action on a function, say, from
$\,L^\infty(\NR^d)\,$ and, in its dependence on $\,\Nv$, \,belongs
to $\,L^p\,$ of the Gaussian process  with $\,p<\infty$. \,The
only problem is the convergence of the series that was established
for $\,\kappa\geq0\,$ ($\kappa=0$ included!) in (Le Jan \& Raimond,
1999). The argument is quite simple. First, for $\,f\in
L^\infty(\NR^d)\,$ and $\,P_{_{\leq N}}(\Nv|t,s)
=\sum\limits_{n=1}^N P_n(\Nv|t,s)\,$ one shows inductively the
{\it a priori} bound: \qq {\bf E}\ |P_{_{\leq N}}(\Nv|t,s)\,f|^2 \
\leq\ P_{_0}(t,s)\,|f|^2\,, \label{ineq} \qqq where the $0^{\rm
th}$-order term $\,P_{_0}(t,s)=\ee^{\,(t-s)\,
\tilde\kappa\Nna^2}$. \,But the terms $\,\,P_n(\Nv|t,s)\,f\,\,$
are orthogonal with respect to the scalar product in $\,L^2\,$
since all the differentials $\,d\NV(\sigma_n)\,$ are independent
in virtue of the It\^{o} convention and their expectations vanish.
\,Hence the bound (\ref{ineq}) establishes the convergence of the
series $\,\,\sum\limits_n P_n(\Nv|t,s)\,f\,\,$ in the $\,L^2\,$
norm of the velocity process for any $\,\kappa\geq0\,$ together
with the limiting bound \qq {\bf E}\ |P(\Nv|t,s)\,f|^2 \ \leq\
P_{_0}(t,s)\,|f|^2\,. \label{lineq} \qqq The continuity of the
result in $\,\kappa\geq0\,$ may be also established. \vskip 0.5cm

\noindent {\bf Proof of estimate} (\ref{ineq}). \ \ For $\,N=0$,
the bound reduces to the easy estimate $\,|P_{_0}(t,s)\,f|^2\leq
P_{_0}(t,s)\,|f|^2$. \,Suppose now that (\ref{ineq}) holds up to
$\,N$. \,Note that \qq P_{_{\leq N+1}}(\Nv|t,s)\,f\ =\
P_{_0}(t,s)\,f\ -\ \int\limits_s^t P_{_{\leq
N}}(\Nv|t,\sigma)\,\,d\NV(\sigma)\cdot\Nna
\,P_{_0}(\sigma,s)\,f\,. \label{blast} \qqq Squaring and taking
expectations, we obtain \qq &&{\bf E}\ |P_{_{\leq
N+1}}(\Nv|t,s)\,f|^2\ \ =\ \ |P_{_0}(t,s)\,f|^2\cr&&+\
\int\limits_s^td\sigma\,\,{\bf E} \,\left(P_{_{\leq
N}}(\Nv|t,\sigma) \otimes P_{_{\leq
N}}(\Nv|t,\sigma)\right)\,{\CK}_2\,\left(P_{_0}(\sigma,s)
\,\ov{f}\,\otimes\,P_{_0}(\sigma,s)\,f\right)\,, \label{last} \qqq
where
$\,{\CK}_2=D^{ij}(\Nx_1-\Nx_2)\nabla_{{x_1^i}}\nabla_{{x_2^j}}\,$
is the operator acting on functions on $\,\NR^d\times\NR^d$.
\,Rewriting the positive-definite function $\,D^{ij}(\Nx)\,$ with
the use of the Fourier transform as $\,\int\hat
D^{ij}(\Nk)\,\ee^{\,i\,\Nk\cdot \Nx}\,d\Nk\,$ with the positive
matrix $\,\hat D^{ij}(\Nk)
=\sum\limits_{\alpha=1}^{d-1}\ov{\lambda^i_\alpha(\Nk)}\,\lambda^j_\alpha
(\Nk)$, \,we may present the second term on the right hand side of
(\ref{last}) as \qq \int\limits_s^td\sigma\int
d\Nk\,\sum\limits_\alpha\ {\bf E}\,\,|P_{_{\leq
N}}(\Nv|t,\sigma)\,f_{\sigma,\Nk,\alpha}|^2\,, \qqq where
$\,\,f_{\sigma,\Nk,\alpha}(\Nx)=\ee^{-i\Nk\cdot\Nx}\,\lambda^i_{\alpha}
(\Nk)\,\nabla_{_{i}}[P_{_0}(\sigma,s)\,f](\Nx)$. \,\,By the
inductive hypothesis, this expression is bounded by \qq
\int\limits_s^td\sigma\int
d\Nk\,\sum\limits_\alpha\,P_{_0}(t,\sigma)
\,|f_{\sigma,\Nk,\alpha}|^2&=&\int\limits_s^td\sigma \int
d\Nk\,\,P_{_0}(t,\sigma)\,\hat
D^{ij}(\Nk)\,\left(\nabla_{_i}P_{_0} (\sigma,s)\,{\ov
f}\right)\left(\nabla_{_j}P_{_0}(\sigma,s)\,f\right)\cr
&&\hspace{-3.5cm}=\
D_0\int\limits_s^td\sigma\,\,P_{_0}(t,\sigma)\, \left|\nabla
P_{_0}(\sigma,s)\,f\right|^2 \ \leq\
2\,\tilde\kappa\int\limits_s^td\sigma\,\,P_{_0}(t,\sigma)\,
\left|\nabla P_{_0}(\sigma,s)\,f\right|^2\,. \qqq Altogether, we
obtain \qq &&{\bf E}\ |P_{_{\leq N+1}}(\Nv|t,s)\,f|^2\ \leq\
|P_{_0}(t,s)\,f|^2\ +\ 2\,\tilde\kappa\int\limits_s^td\sigma\,
P_{0}(t,\sigma)\,\left| \nabla P_{_0}(\sigma,s)\,f\right|^2\cr
&&=\ |P_{_0}(t,s)\,f|^2\ -\ \int\limits_s^td\sigma\,\,{d\over
d\sigma} \left(P_{_0}(t,\sigma)\,\,|P_{_0}(\sigma,s)\,f|^2\right)
\ =\ P_{_0}(t,s)\,|f|^2 \qqq ending the inductive proof of
eq.\,\,(\ref{ineq}). \ $\Box$ \vskip 0.4cm

The Chapmann-Kolmogorov chain relation
$\,\,P(\Nv|t,\sigma)\,P(\Nv|\sigma,s)= P(\Nv|t,s)\,\,$ for
$\,s\leq\sigma\leq t\,$ and the normalization
$\,\,P(\Nv|t,s)\,1=1\,\,$ follow easily. \,Le Jan and Raimond also
proved that the operators $\,P(\Nv|t,s)\,$ preserve positivity.
This implies that \qq |P(\Nv|t,s)\,f|\ \leq\ P(\Nv|t,s)\,|f|\
\leq\ P(\Nv|t,s)\,\Vert f \Vert_{_\infty}\ =\ \Vert
f\Vert_{_\infty} \qqq so that $\,P(\Nv|t,s)\,f\,$ are
(essentially) bounded. \,We obtain this way a family of Markov
transition probabilities parametrized by the velocities of the
Kraichnan ensemble, hence, also a family of Markov processes
describing the Lagrangian trajectories (with and without the
perturbing noise). Taking the $\,N\to\infty\,$ limit in
eq.\,\,(\ref{blast}), we infer that $\,P(\Nv|t,s)\,$ satisfies the
stochastic (anti-)It\^{o} integral equation \qq P(\Nv|t,s)\,f\ =\
P_{_0}(t,s)\,f\ -\ \int\limits_s^t
P(\Nv|t,\sigma)\,\,d\NV(\sigma)\cdot\Nna \,P_{_0}(\sigma,s)\,f\,.
\label{blast1} \qqq which, \,upon differentiation over $\,s$,
\,gives the stochastic versions of the diffusion-advection
equation \qq d_s\,P(\Nv|t,s)\,f&=&P(\Nv|t,s)\,d\NV(s)\cdot\Nna f
-\tilde\kappa\,P(\Nv|t,s)\,\Nna^2f\,ds\label{sad0}\\\cr
&=&P(\Nv|t,s)\circ d\NV(s)\cdot\Nna f
-\kappa\,P(\Nv|t,s)\,\Nna^2f\,ds\,. \label{sad1} \qqq \vskip 0.8cm

\noindent 2.3. \ {\bf $N$-particle processes in Kraichnan velocities}
\vskip 0.5cm

\noindent As follows for example from the relation (\ref{blast1}),
the expectation of the transition probability $\,P(\Nv|t,s)\,$
coincides with the first term $\,P_{_0}(t,s)=\ee^{\,(t-s)\,\tilde
\kappa\Nna^2}$ of the series (\ref{ito}). Thus, recalling the
definition (\ref{san}), we obtain \qq \CP_{_1}(t,\Nx;s,d\Ny)\ =\
\ee^{\,(t-s)\,\tilde\kappa\Nna^2}(\Nx,\Ny) \,d\Ny\,. \qqq We infer
that in the Kraichnan model, a single fluid particle undergoes
diffusion for all times with the diffusion constant equal to
$\,\tilde\kappa$, \,i.e.\,\,to the sum of the molecular
diffusivity $\,\kappa\,$ and of the ``eddy diffusivity'' $\,\hf
D_0$, an effective diffusivity due to the random velocities.
Recall that $\,D_0=\CO(L^\xi)\,$ so that the eddy diffusivity is
dominated by the integral scale, i.e. by the correlation length of
the velocities. The virtue of the use of the stochastic It\^{o}
integrals in (\ref{ito}) was that it made the regularizing role of
the eddy diffusion explicit and permitted uniform treatment of the
cases with $\,\kappa>0\,$ and with $\,\kappa=0$. \vskip 0.3cm

In order to study the statistics of $\,N\,$ particles in the
Kraichnan model, we have to analyze the joint transition
probabilities $\,\CP_{_N}(\un{t},\un{\Nx};\un{s},d\un{\Ny})$,
\,see (\ref{san}). If fact, it is enough to look at their
equal-time versions $\,\CP_{_N}({t},\un{\Nx};{s},d\un{\Ny})$.
\,The stochastic (anti-)It\^{o} equation (\ref{sad0}) implies the
relation \qq
{_d\over^{ds}}\int\CP_{_N}(t,\un{\Nx};s,d\un{\Ny})\,f(\un{\Ny})&=&
-\sum\limits_{n<m}\int\CP_{_N}(t,\un{\Nx};s,d\un{\Ny})
\,\,D^{ij}(\Ny_n-\Ny_m)\,\nabla_{y_n^i}\nabla_{y_m^j}f(\un{\Ny})\cr
&&-\,\sum\limits_n\int\CP_{_N}(t,\un{\Nx};s,d\un{\Ny})\,\tilde\kappa
\Nna_{\Ny_n}^2 f(\un{\Ny}) \qqq from which one deduces that \qq
\CP_{_N}(t,\un{\Nx};s,d\un{\Ny})\ =\
\ee^{\,(t-s)\,\CM_{_N}}(\un{\Nx}, \un{\Ny})\,\,d\un{\Ny}
\label{hk} \qqq for the second order differential operator \qq
\CM_{_N}\ =\ \hf\sum\limits_{n,m}D^{ij}(\Nx_n-\Nx_m)\nabla_{x_n^i}
\nabla_{x_m^j}\ +\ \kappa\sum\limits_n\nabla_{x_n}^2\,. \label{mn}
\qqq It follows that in the Kraichnan model the transition
probabilities $\,\CP_{_N}(t,\un{\Nx};s,d\un{\Ny})\,$ still form a
Markov family (this is due to the temporal decorrelation of
velocities). The probabilities of relative separations of $\,N\,$
particles \qq \CP^{\rm rel}_{_N}(t,\un\Nx;s,d\un\Ny)\ =\ \int
\CP_{_N}(t,\un\Nx;s,d(\Ny_1+\Ny),\dots,d(\Ny_{_N}+\Ny)\,, \qqq with
the integral over the translations $\,\Ny$, \,are given by the
exponential function of the operators $\,M_{_N}\,$ that are
restrictions of $\,\CM_{_N}$'s \,to the translation-invariant
sector: \qq M_{_N}\ =\
-\sum\limits_{n<m}\left(d^{ij}(\Nx_n-\Nx_m)+2\kappa\right)
\nabla_{x_n^i}\nabla_{x^j_m} \label{mnr} \qqq with
$\,d^{ij}(\Nx)\,$ given by (\ref{dd}). As we see, in the Kraichnan
velocities, $\,N\,$ fluid particles undergo in their relative
motion an effective diffusion process with the
configuration-dependent diffusivity. \vskip 0.3cm

In particular for the probability distribution (\ref{rel}) of the
relative separation of two fluid particles, we obtain \qq \CP^{\rm
rel}_{_2}(t,\bm\rho_0;s,d\bm\rho)\ =\
\ee^{\,(t-s)\,M_2}(\bm\rho_0, \bm\rho) \,d\bm\rho\,,\label{2rel}
\qqq where \qq M_2\ =\
d^{ij}(\bm\rho)\,\nabla_{i}\nabla_{j}\,+\,2\,\kappa\Nna^2\,.
\label{m2} \qqq This operator commutes with the action of the
rotation group in $\,L^2(\NR^d)\,$ and reduces in the action on
functions carrying irreducible representations of $\,SO(d)\,$
(labeled by the angular momentum $\,\ell=0,1,\dots$) to a second
order differential operator in the radial variable
$\,\rho=|\bm\rho|$. In particular, the probability distribution
$\,\CP_{_2}(t,\rho_0;s,d\rho)\,$ of the 2-particle
dispersion is given by the exponential function of the restriction
$\,M_2^{{\ell=0}}\,$ of operator $\,M_2\,$ to the
rotation-invariant sector with $\ell=0$. \vskip 0.3cm

The above relations will allow us to analyze in the next two
lectures the properties of the fluid particles in the Kraichnan
ensemble of velocities. \vskip 1cm

\nsection{3}
\vskip 0.3cm

\noindent ({\small fluid particles and advection of scalar in the Batchelor
regime  of the Kraichnan model: random chaos})
\vskip 0.8cm

\noindent 3.1. \ {\bf Separation of close particles in smooth
Kraichnan velocities}
\vskip 0.5cm

\noindent In the region where the distances between the fluid
particles are much smaller then the viscous scale $\,\eta$,
\,i.e.\,\,in the Batchelor regime, we may approximate $\,d^{ij}\,$
as in (\ref{smr}). Upon dropping the $4^{\rm th}$-order terms
(which, strictly speaking pertains to the behavior of
infinitesimally close trajectories) and upon setting $\,\kappa=0$,
\,the generators of the $\,N$-particle processes become \qq
M_{_N}\ =\
-D_1\sum\limits_{n<m}\left((\Nx_n-\Nx_m)^2\Nna_{\Nx_n}\cdot
\Nna_{\Nx_m}\ +\ 2\,(x_n^i-x_m^i)(x_n^i-x_m^j)\nabla_{x_n^i}
\nabla_{x_m^j}\right)\,. \label{mns} \qqq In particular, for
$\,N=2$, \qq M_2\ =\
D_1\,[\,\bm\rho^2\Nna^2\,-\,2\,\rho^i\rho^j\nabla_i\nabla_j] \qqq
in terms of the separation variable $\,\bm\rho$. In the rotational
invariant sector, \qq M_2^{{\ell=0}}\ =\
D_1(d-1)\,\rho^{-d+1}\partial_\rho\,\rho^{d+1}
\partial_\rho\,.
\label{roti}
\qqq
 From the latter expression, one infers easily the probability
distribution of the 2-particle dispersion:
\qq
\CP_{_2}(0,\rho_0;t,d\rho)\ =\ {_1\over^{\sqrt{4\pi\,D_1(d-1)\,t}}}
\,\,\exp\Big[-{_1\over^{4\,D_1(d-1)\,t}}\left(\ln({_\rho\over^{\rho_0}})
-D_1(d-1)d\,t\right)^2\Big]\,\,{_{d\rho}\over^{\rho}}\,.
\label{lyap}
\qqq
As we see, the logarithm of the 2-particle dispersion
grows linearly with the rate (the Lyapunov exponent)
$\,\lambda=D_1(d-1)d>0$. The system exhibits the exponential
separation of trajectories, i.e.\,\,is chaotic.
It is easy to see that
\qq
\lim\limits_{\rho_0\to0}\ \CP_{_2}(0,\rho_0;t,d\rho)\ =\ \delta(\rho)\,
d\rho\,,\label{Ba}
\qqq
which is a consequence of the existence of deterministic Lagrangian
trajectories in smooth velocities: it signals that the Markov
process with the transition probabilities $\,P(\Nv|t,\Nx;s,d\Ny)\,$
in a fixed velocity $\,\Nv\,$ realization concentrates (for $\,\kappa=0$) on
deterministic trajectories determined by the initial conditions.
\vskip 0.3cm

It is not difficult to understand the origin of the positivity of
the Lyapunov exponent in the smooth Kraichnan velocities. For such
velocities, the equation (\ref{sep}) for very close trajectory
separation may be approximated by \qq {d{\bm\rho}\over ds}\ =\
{\bm\rho}\cdot\Nna\Nv(s,{\bm x}_{0,\Nx_1}(s))\,. \label{prec} \qqq
In the Kraichnan model, \qq {\bf E}\
\nabla_{_k}v^i(t,\Nx)\,\nabla_{_\ell}v^j(s,\Ny)\ =\
\delta(t-s)\,\,\nabla_{_k}\nabla_{_\ell}\,d^{ij}(\Nx-\Ny) \qqq
which is independent of $\,(\Nx-\Ny)\,$ in the quadratic
approximation to $\,d^{ij}(\Nx)$. \,In other words, in this
approximation, $\,\nabla_{_k}v^i(t,\Nx)=\gamma^{ik}(t)\,$ where
$\,\gamma^{ik}(t)\,$ is a (Gaussian) white noise with the values
in the real traceless matrices, with mean zero and covariance \qq
{\bm E}\ \gamma^{ik}(t)\,\gamma^{j\ell}(s)\ =\
2\,D_1\,\delta(t-s)\,\Big(
(d+1)\,\delta^{ij}\,\delta^{kl}\,-\,\delta^{ik}\,\delta^{j\ell}
\,-\,\delta^{i\ell}\,\delta^{jk}\Big)\,. \qqq Eq.\,\,(\ref{prec})
becomes the stochastic differential equation \qq d{\bm\rho}\ =\
d\Gamma\circ{\bm\rho} \qqq where $\,\Gamma(s)\,$ is a Brownian
motion on traceless matrices with $\,d\Gamma(s)=\gamma(s)\,ds$.
\,The solution is given by \qq {\bm\rho}(s)\ =\
G_t(s)\,{\bm\rho}(t)\,, \qqq where $\,G_t(s)\,$ is a Brownian
motion on the group $\,SL(d)\,$ of real unimodular matrices
satisfying $\,G_t(t)=1$. \,It follows that in the quadratic
approximation to $\,d^{ij}(\Nx)$, \qq \int
P(\Nv|t,\Nx_1;s,d\Ny_1)\,\,P(\Nv|t,\Nx_2;s,d\Ny_2)\,\,f(\Ny_2-\Ny_1)
\ =\ f\left(G_t(s)\bm\rho_0\right)\,, \label{bav} \qqq where
$\,\rho_0=\Nx_2-\Nx_1$. \,Taking the averages of the last
identity, we infer that \qq \Big(\CP_{_2}^{\rm
rel}(t,s)\,f\Big)(\bm\rho_0)\ =\ {\bf E}\
f\left(G_t(s)\bm\rho_0\right)\,. \qqq It is not difficult to find
the generator of the Brownian motion $\,G_t(s)$. \,Consider the
natural actions of $\,SL(d)\,$ and of its subgroup $\,SO(d)\,$ in
$\,L^2(\NR^d)$. Their infinitesimal generators are \qq &&H^{ij}\
=\ \left(-\rho^i\nabla_{j}\,+\,{_1\over^d}\,
\delta^{ij}\,\rho^k\nabla_{k}\right)\,,\cr &&J^{ij}\ =\
\left(-\rho^i\nabla_{j}+ \rho^j\nabla_{i}\right)\,, \qqq
respectively. A simple algebra shows that \qq M_{2}\ =\
D_1\,[\,d\,H^2\,-\,(d+1)\,J^2]\, \label{cas2} \qqq where $\,H^2\,$
and $\,J^2\,$ are the quadratic Casimirs of $\,SL(d)\,$ and of
$\,SO(d)\,$: \qq H^2\ =\ H^{ij}\,H^{ji}\,,\qquad J^2\ =\ -\hf
(J^{ij})^2\,. \label{cas} \qqq The right hand side of (\ref{cas2})
(with the Casimirs interpreted as those of the left regular
action) gives the generator of the Brownian motion $\,G_t(s)\,$ on
$\,SL(d)$. \,It is well known that such Brownian motions, that may
be thought of as  continuous products of random, independent,
identically distributed matrices in $\,SL(d)$, \,lead to positive
Lyapunov exponents (a continuous version of the Furstenberg-Kesten
or Oseledets Theorems, see e.g.\,\,(Arnold, 1998)). \vskip 0.3cm

For the $\,N$-point operator $\,M_{_N}\,$ given by (\ref{mns}), we
similarly obtain: \qq M_{_N}\ =\
D_1\,[\,d\,H_{_{N-1}}^2\,-\,(d+1)\,J_{_{N-1}}^2]\, \qqq where
$\,H_{_{N-1}}^2\,$ and $\,J_{_{N-1}}^2\,$ are the quadratic
Casimirs of the diagonal actions of $\,SL(d)\,$ and of $\,SO(d)\,$
on the $\,(N-1)\,$ separation variables, e.g.\,\,on
$\,\bm\rho_n=\Nx_n-\Nx_{1}$. \,An alternative expression for
$\,M_{_N}\,$ is \qq M_{_N}\ =\
D_1\,[\,d\,G_{_{N-1}}^2\,+\,{_{d-N+1}\over^{N-1}}
\,\Lambda(\Lambda+(N-1)d)\,-\,(d+1)\,J_{_{N-1}}^2]\, \label{sln1}
\qqq where $\,G^2_{_{N-1}}\,$ is the quadratic Casimir of the
action of $\,SL(N-1)\,$ on the index $\,n\,$ of $\,\rho^i_{n}\,$
with the generators $\,\,G^{nm}=
-\rho^i_{n}\nabla_{\rho^i_{m}}+{1\over N-1}\,\delta^{nm}\,
\sum\limits_p x^i_{p}\nabla_{x_{p}^i}\,\,$ and
$\,\,\Lambda=\sum\limits_p x^i_{p}\nabla_{x^i_{p}}\,\,$ is the
generator of the overall dilations. Consider a function $\,f\,$ of
$\,N-1\,$ separations that depends only on the volume
$\,\,\rho=\sqrt{{\rm det}_{_{nm}}(\bm\rho_{n}
\cdot\bm\rho_{m})}\,\,$ spanned by $\,N-1\,$ separation vectors
(for $\,N-1\leq d$). \,Since $\,\rho\,$ is $\,SL(N-1)\,$ \,and
$\,SO(d)$-invariant, it follows from (\ref{sln1}) that
$\,M_{_N}f\,$ is still a function of $\,\rho\,$ only and that \qq
\left(M_{_N}f\right)(\rho)&=&D_1\,{_{d-N+1}\over^{N-1}}\,
\Lambda(\Lambda+(N-1)d)\,f(\rho)\cr\cr
&=&D_1\,(N-1)(d-N+1)\,(\rho^{-d+1}\,\partial_{\rho}
\rho^{d+1}\,\partial_{\rho}\,f(\rho)\,. \qqq Comparing to
(\ref{roti}) and (\ref{lyap}), we infer that the logarithm of
$\,\rho\,$ grows linearly with the rate $\,D_1(N-1)(d-N+1)d$. By
definition, the latter gives the sum
$\,\lambda_1+\dots+\lambda_{_{N-1}}\,$ of the $\,(N-1)\,$ biggest
Lyapunov exponents so that \qq \lambda_N\ =\ D_1(d-2N+1)d\,. \qqq
In the smooth $\,d$-dimensional Kraichnan flow the $\,d\,$
Lyapunov exponents are equidistant, with the sum equal to zero (as
required by incompressibility). \,The relations of the
multi-trajectory processes to the harmonic analysis on the groups
$\,SL(N)\,$ were observed in (Shraiman \& Siggia, 1995 and 1996),
\vskip 0.7cm

\noindent 3.2 \ {\bf Scalar statistics in the Batchelor regime}
\vskip 0.5cm

\noindent One may easily control the stationary state of the
scalar developing under the forced transport of Kraichnan
velocities in the Batchelor regime. For white in time velocities,
in agreement with the preceding discussion, the solution of the
forced transport problem is given by eq.\,\,(\ref{han}) with the
evolution operators $\,P(\Nv|t,s)\,$ as constructed above. For
simplicity, let us take the zero initial data for the scalar:
$\,\theta(s)=0$. \,Then the characteristic function of the time
$\,t\,$ probability distribution of the scalar, \qq \Phi(h|t,s)\
=\ {\bm E}\ \ee^{\,i\int d\Nx\,\, h(\Nx)\,\theta(t,\Nx)} \ =\ {\bm
E}\ \ee^{\,i\int\limits_s^td\sigma\int d\Nx\,\,
h(\Nx)\,\,P(\Nv|t,\Nx;\,\sigma,d\Ny)\,\,f(\sigma,\Ny)}\,.
\label{fsr} \qqq Let us take the source $\,f(s,\Ny)\,$ to be also
a Gaussian process, independent of velocities, with mean zero and
covariance \qq {\bf E}\ f(t,\Nx)\,f(s,\Ny)\ =\
\delta(t-s)\,\,\chi(\Nx-\Ny)\,,\label{fsr1}\qqq where $\,\chi\,$
is a smooth, positive-definite, decaying function on $\,\NR^d$.
For such a forcing, the expectation over the source $\,f\,$ of the
scalar in (\ref{fsr}) may be easily calculated, leaving us with
the expectation over the velocity ensemble: \qq \Phi(h|t,s)\ =\
{\bm E}\ \ee^{-\hf\int\limits_s^td\sigma \int
d\Nx_1\,d\Nx_2\,\,h(\Nx_1)\,h(\Nx_2)\,\,
P(\Nv|t,\Nx_1;\,\sigma,d\Ny_1)\,\,P(\Nv|t,\Nx_2;\,\sigma,d\Ny_2)\,\,
\chi(\Ny_2-\Ny_1)}\,. \qqq In the Batchelor regime, we may use the
relation (\ref{bav}) to rewrite the latter expectation in terms of
the one over the Brownian motion $\,G_t(\sigma)\,$ on $\,SL(d)$:
\qq \Phi(h|t,s)\ =\ {\bm E}\ \ee^{-\hf\int\limits_s^td\sigma \int
d\Nx_1\,d\Nx_2\,\,h(\Nx_1)\,h(\Nx_2)\,\,\chi\left(G_t
(\sigma)(\Nx_1-\Nx_2)\right)}\ =\ {\bm E}\ \ee^{-\int\limits_s^t
V_h\left(G_t(\sigma)\right)\,\,d\sigma}\,,\label{Vh} \qqq where
$\,\,V_h(G)=\hf\int\int d\Nx_1\,d\Nx_2\,\,h(\Nx_1)\,h(\Nx_2)
\,\chi\left(G(\Nx_1-\Nx_2)\right)\,\,$ is a positive potential on
group $SL(d)$. \,With the use of the Feynman-Kac formula,
eq.\,\,(\ref{Vh}) may be expressed as \qq
\Phi(h|t,s)&=&\int\limits_{SL(d)}
\ee^{\,(t-s)\,[d\,H^2-(d+1)\,J^2-V_h]}(1,G)\,\,dG\cr
&=&1-\int_0^{t-s}d\sigma\
\Big(\ee^{\,\sigma\,[d\,H^2-(d+1)\,J^2-V_h]}\,V_h \Big)(1)\,, \qqq
where the last equality follows by integration by parts. It is
then easy to show that the limit
$\,\,\Phi(h)=\lim\limits_{s\to-\infty}\,\Phi(h|t,s)\,\,$ exists
and is a characteristic function of a probability measure
supported e.g.\,\,on the space $\,\CS'(\NR^d)\,$ of tempered
distributions. The latter describes the stationary state of the
forced scalar which, indeed, is not supported by smooth scalar
configurations, as reflected by logarithmic singularities at
coinciding points of the scalar $N$-point functions. For
$\,\chi\,$ and $\,h\,$ rotation-invariant, potential $\,V_h\,$ is
left and right $\,SO(d)$-invariant. On $\,SO(d)\backslash
SL(d)/SO(d)$, \,the operator $\,[d\,H^2-(d+1)\,J^2]\,$ reduces the
the Calogero-Sutherland integrable Hamiltonian and the analysis of
the distribution of the random variable $\,\int d\Nx\,h(\Nx)
\,\theta(\Nx)\equiv\theta(h)\,$ reduces to the analysis of the
property of its perturbation by the potential proportional to
$\,V_h$. \,For example, the exponential rate of decay of the
probability density function of $\,\theta(h)\,$ is related to the
bound state energy of such a perturbation with a negative
coefficient. The details may be found in (Bernard, Gawedzki \&
Kupiainen, 1998). \vskip 1cm

\nsection{4}
\vskip 0.3cm

\noindent ({\small fluid particles in non-smooth Kraichnan
velocity fields; breakdown of deterministic Lagrangian flow;
intrinsic stochasticity versus particle aggregation})
\vskip 0.8cm

\noindent 4.1. \ {Separation of close particles in non-smooth
Kraichnan velocities}
\vskip 0.5cm

\noindent In the limiting case $\,\eta\to0\,$ of the Kraichnan
model, the Lagrangian trajectories exhibit even more dramatic
behavior. For $\,\eta=0\,$ and $\,L=0$, \,when $\,d^{ij}(\Nx)\,$
takes the scaling form of eq.\,\,(\ref{lsf}), and for
$\,\kappa=0$, \qq M_2^{{\ell=0}}\ =\
D_2(d-1)\,\rho^{-d+1}\partial_\rho\,\rho^{d-1+\xi}
\partial_\rho\,.
\label{roti1} \qqq The probability distribution of the 2-particle
dispersion \qq \CP_{_2}(0,\rho_0;t,d\rho)\ =\
\ee^{\,t\,M_2^{\ell=0}}(\rho_0,\rho)\,d\rho \qqq may be studied
for example using the spectral decomposition of $\,M_2^{\ell=0}$.
The long-time-large-distance asymptotics of the dispersion follows
from the rescaling property \qq \CP_{_2}(0,\rho_0;\mu
t,d(\mu^{1/(2-\xi)}\rho))\ = \
\CP_{_2}(0,\mu^{-1/(2-\xi)}\rho_0;t,d\rho) \qqq and the easily
shown relation \qq \lim_{\rho_0\to0}\ \CP_2(0,\rho_0;t,d\rho)\
\propto\ {\rho^{d-1}}\,t^{-d/(2-\xi)}\,\,\ee^{-{\rm
const}.\,\rho^{2-\xi}/t}\,\, d\rho\,.\label{ll} \qqq In
particular, one obtains the Richardson dispersion law in the form
\qq {\bf E}\ \rho(t)^2\ =\ \int\rho^2\,\,
\CP_{_2}(0,\rho_0;t,d\rho)\ = \ \CO(t^{^{2\over2-\xi}}) \non \qqq
for large times, with an exact proportionality for all times in
the limit $\,\rho_0\to0$. \,Such a behavior may be predicted by
solving the modified version  \qq d\rho^2\ \propto\
dw\,\,\rho^{\xi/2\,+\,1}\qqq of the naive mean field type equation
(\ref{alp0}), where $\,w(t)\,$ is the Brownian motion introduced
to account for the temporal decorrelation of Kraichnan velocities,
In particular, the growth of $\,\rho^2\,$ proportional to
$\,t^3\,$ is obtained for $\,\xi=4/3$. \vskip 0.3cm

Note that the limit (\ref{ll}) is absolutely continuous with
respect to the Lebesque measure $\,d\rho$, in contrast to what
happens in the Batchelor regime, see eq.\,\,(\ref{Ba}). Such a
property excludes the concentration of the transition
probabilities $\,P(\Nv|t,\Nx;s,d\Ny)\,$ at a single
($\Nv$-dependent) point $\Ny$ and signifies the breakdown of the
deterministic Lagrangian flow with fluid particle trajectories
determined by their initial positions in fixed velocity fields.
Instead, the Lagrangian trajectories form a genuinely stochastic
process already in a fixed velocity realization, as predicted
above and illustrated in the figure below:

\leavevmode\epsffile[25 15 400 400]{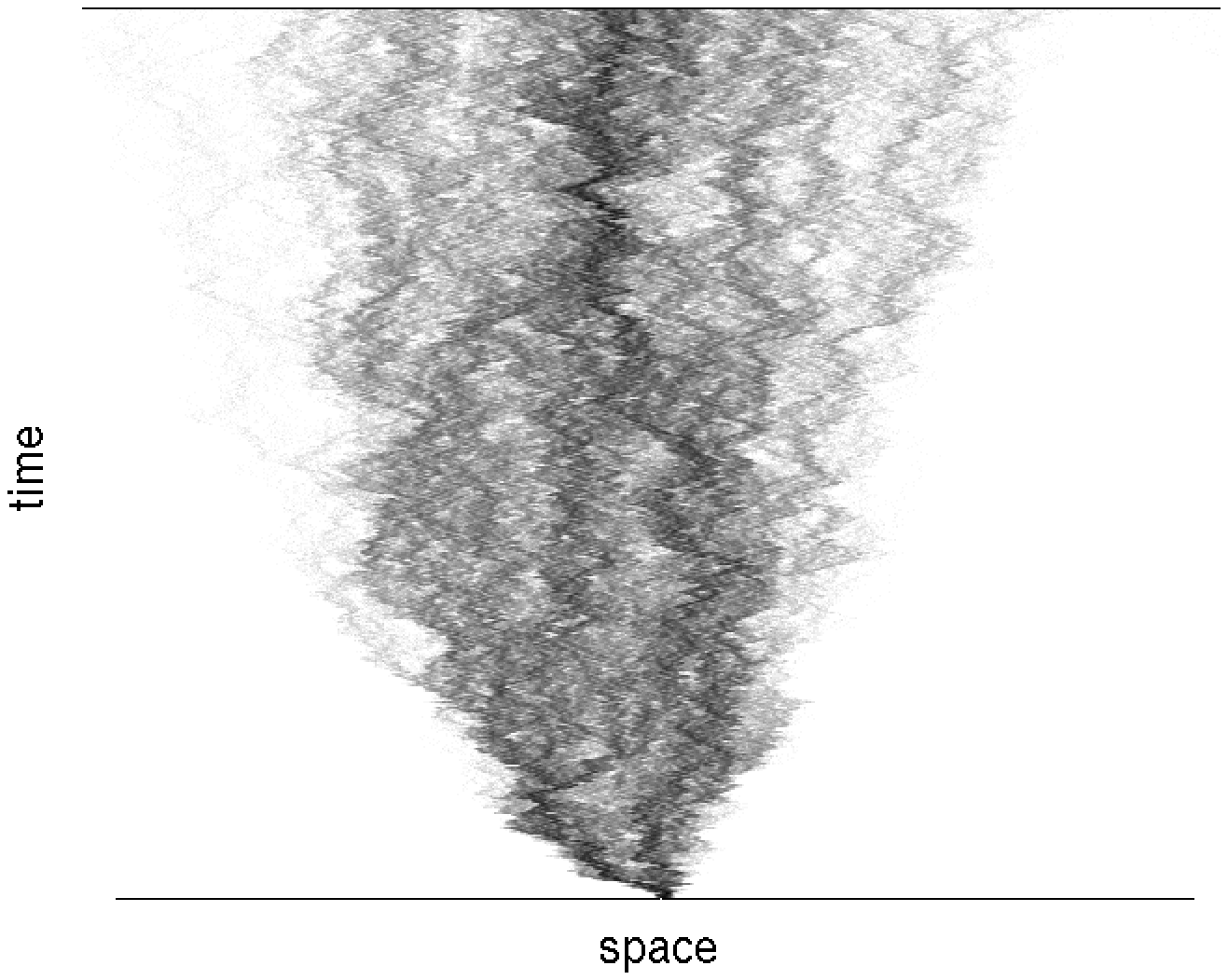}

\noindent That Markov process, may be obtained by first adding the
regularizing noise to the trajectory equation, as in (\ref{sode}),
and then turning it off, or also directly, by the Le Jan-Raimond
construction of the $\,\kappa=0\,$ transition probabilities
$\,P(\Nv|t,\Nx;s,d\Ny)$. \vskip 0.7cm

\noindent 4.2. \ {\bf The role of compressibility} \vskip 0.5cm

\noindent Up to now, we have considered the Kraichnan ensemble with
incompressible velocities, the property guaranteed by the presence
of the transverse projector $\,(\delta^{ij}-{k^ik^j\over k^2})\,$
in the Fourier representation (\ref{dd0}) of the velocity 2-point
function. In order to study the effects of compressibility, one
may introduce another parameter besides the roughness exponent
$\,\xi$, \,the compressibility degree $\,\wp$. \,We shall do
it by replacing the transverse projector in (\ref{dd0}) by
$\,[(1-\wp)\delta^{ij}+(\wp d-1){k^ik^j\over k^2}]$. \,The value
$\,\wp=0\,$ corresponds to
the incompressible case whereas for $\,\wp=1\,$ almost all
velocities are gradients, with the intermediate values of $\,\wp\,$
interpolating between the two cases. The preceding constructions,
in particular the Le Jan-Raimond one, carry over to the case with
non-zero $\,\wp$. \,The scaling form of the generator of the
2-particle dispersion process becomes now \qq M_2^{{\ell=0}}\ =\
D_2(d-1)\,\rho^{\xi-a}\partial_\rho\,\rho^a
\partial_\rho
\label{rotip} \qqq with $\,a={d+\xi\over1+\wp\xi}-1$. \,The
definition of the semigroup $\,\ee^{\,t\,M_2^{\ell=0}}\,$ requires
a choice of the boundary condition for the operator
$\,M_2^{\ell=0}\,$ at $\,\rho=0$. \,Such a choice is automatically
assured by considering first the $\,\kappa>0\,$ case and then
sending $\,\kappa\,$ to zero. This limiting procedure selects for
$\,\wp<d/\xi^2$, \,i.e.\,\,for weak compressibility, the
eigenfunctions of $\,M_2^{\ell=0}$ behaving like $\CO(1)$ at
$\rho=0$ whereas for strong compressibility $\,\wp>d/\xi^2\,$ the
eigenfunctions that behave like $\,\CO(\rho^{1-a})$ are chosen.
The two choices result in very different probability distributions
$\,\CP_{_2}(0,\rho_0;t,d\rho)\,$ of the 2-particle dispersion and,
in consequence, to dramatically different Lagrangian flows, as
first noticed in (Gawedzki \& Vergassola, 2000). For
$\,\wp<d/\xi^2$, \,in a simple generalization of (\ref{ll}), \qq
\lim_{\rho_0\to0}\ \CP_2(0,\rho_0;t,d\rho)\ \propto\
{\rho^{a-\xi}}\,t^{(\xi-1-a)/(2-\xi)}\,\,\ee^{-{\rm
const}.\,\rho^{2-\xi}/t}\,\, d\rho\,,\label{l2} \qqq  from which
the conclusions about the the intrinsic stochasticity of the
Lagrangian flow may be drawn the same way as for the
incompressible case. For $\,\wp>\xi/d^2$, however, \qq
\CP_{_2}(0,\rho_0;t,d\rho)\ =\ \CP^{\rm
reg}(0,\rho_0;t,d\rho)\,+\,p(t,\rho_0)\,\delta(\rho)\,d\rho\qqq
with $\,\CP^{\rm reg}(0,\rho_0;t,d\rho)\,$ absolutely continuous
with respect to the Lebesque measure $\,d\rho\,$ and
$\,p(t,\rho_0)>0$. When $\,\rho_0\to0\,$ the regular part tends to
zero and $\,p(t,\rho_0)\,$ tends to $1$ so that one recovers the
behavior (\ref{Ba}). The latter indicates that the Lagrangian flow
is deterministic, with trajectories determined by the initial
condition in fixed velocity realizations. The presence of the term
proportional to the delta-function for $\,\rho_0>0\,$ signals,
however, that, with positive $t$-dependent probability, the
trajectories starting at different initial points collapse
together by time $\,t$, \,as illustrated in the figure on the next
page. Such a non-conventional behavior of trajectories is again
possible since the theorem about the existence and unicity of
trajectories fails for non-Lipschitz velocities. The competition
between the tendency of the trajectories in such velocities to
separate explosively and their trapping in the regions of strong
compression is won by the first trend for weak compressibility and
by the second one for the strong one. \vskip 0.7cm

\noindent 4.3. \ {\bf Persistence of scalar dissipation} \vskip
0.5cm

\noindent The unusual behaviors of Lagrangian trajectories in rough
Kraichnan velocities are the source of phenomena crucial for the
scalar transport. One of the implications of the
advection-diffusion equation (\ref{ps}) in the incompressible
velocities is the scalar "energy" $\,\int\theta^2\,$ balance \qq
{_d\over^{dt}}\int\theta(t,\Nx)^2\,d\Nx =\
-2\kappa\int(\bm\nabla\theta(t,\Nx))^2\,d\Nx\,
+\,2\int\theta(t,\Nx)\,f(t,\Nx)\,d\Nx \qqq with the first term on
the right hand describing the dissipation and the second one the
injection of scalar energy by the sources. One may naively expect
that in the absence of sources the scalar energy is conserved in
the limit $\,\kappa\to0$. \,Let us examine this question more
closely. Recall that the evolution of scalar is related by
eq.\,\,(\ref{han0}) to the transition probabilities of the (noisy)
Lagrangian trajectories. For vanishing sources, one obtains the
identity \qq \int d\Nx\int
P(\Nv|t,\Nx;s,d\Ny)\,[\theta(s,\Ny)-\theta(t,\Nx)]^2 \ =\
\int\theta(s,\Ny)^2\,d\Ny\,-\,\int\theta(t,\Nx)^2\,d\Nx \qqq which
follows by developing the square and using eq.\,\,(\ref{han0}),
the normalization of the measure $\,P(\Nv|t,\Nx;s,d\Ny)\,$ and the
symmetry of the operator $\,P(\Nv|t,s)$. \,The left hand side is
obviously non-negative which implies that the scalar energy cannot
grow. It vanishes if and only if $\,\theta(s,\Ny)=\theta(t,\Nx)\,$
on the support of $\,P(\Nv|t,\Nx;s,\Ny)\,$ for each $\Nx$. This
happens for arbitrary $\,\theta(s,\Ny)\,$ if and only if
$\,P(\Nv|t,\Nx;s,d\Ny)\,$ is supported by exactly one
($\Nx$-dependent) point, i.e.\,\,if the Lagrangian trajectories
are determined by their single-time positions. The intrinsic
stochasticity

\leavevmode\epsffile[25 40 400 495]{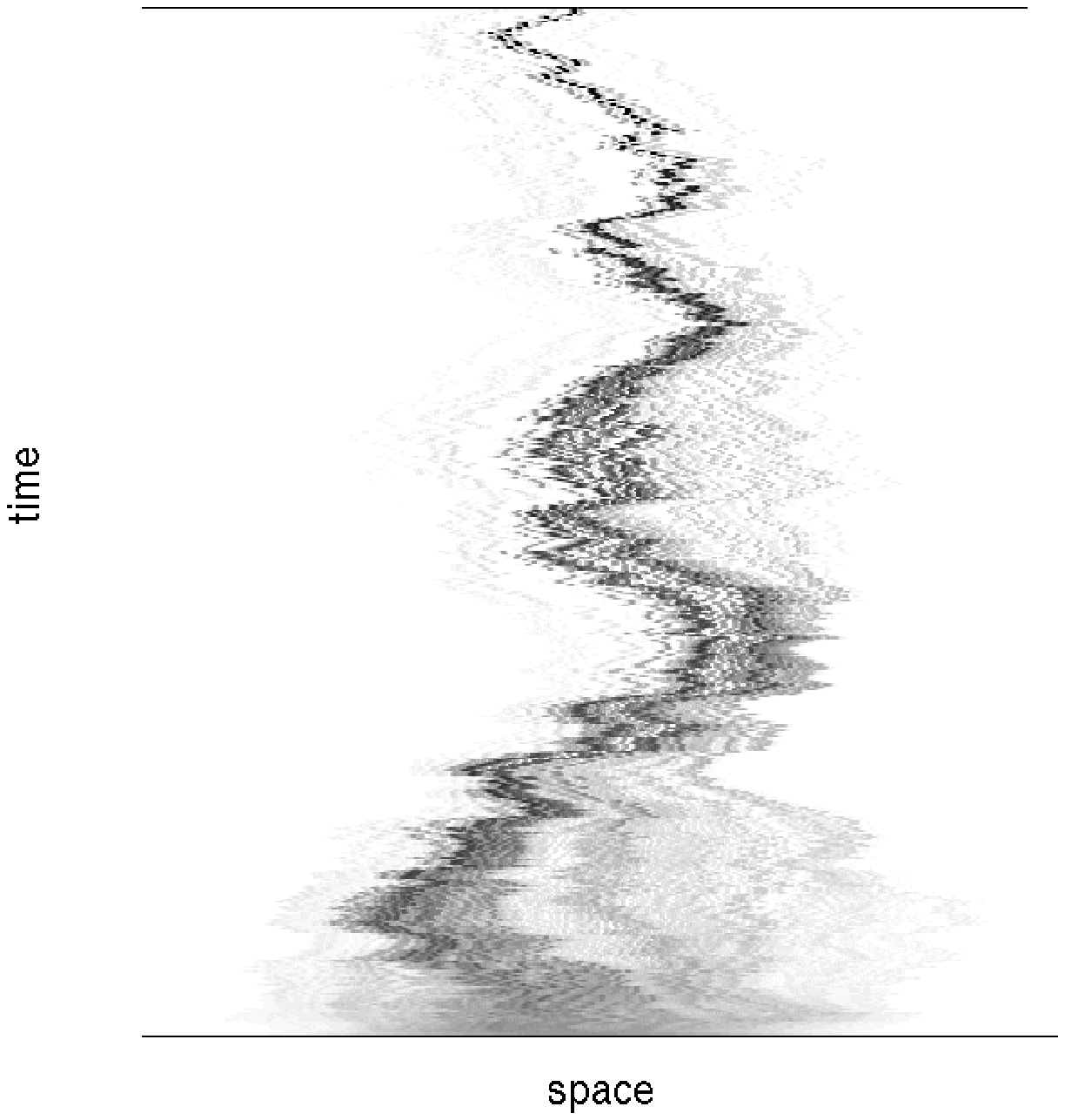}

\noindent of the Lagrangian flow results then in the persistence
of scalar energy dissipation in the limit $\,\kappa\to0$.
\vskip 0.3cm

The same effect may be seen in averaged quantities, also in
compressible Kraichnan velocities. Suppose that at the initial
time $\,s\,$ we are given an homogeneous, isotropic scalar
distribution with the 2-point function \qq {\bf E}\
\theta(s,\Ny)\,\theta(s,\Ny+\bm\rho)\ =\ F_{_2}(s,\rho)\,. \qqq
Then, if the initial scalar distribution is independent of
velocities and in the absence of sources, the scalar 2-point
function at the later time $\,t\,$ is given by \qq
F_{_2}(t,\rho_0)\ \equiv\ {\bf E}\
\theta(t,\Nx)\,\theta(t,\Nx+\bm\rho_0) \ =\ \int
F_{_2}(s,\rho)\,\,\CP_{_2}(t,\rho_0;s,d\rho)\,.\label{frd}\qqq In
particular, the mean scalar energy density \qq F_{_2}(t,0)\ =\
\int F_{_2}(s,\rho)\
\lim_{\rho_0\to0}\,\,\CP_{_2}(t,\rho_0;s,d\rho)\,. \qqq Since
$\,F_{_2}(s,\rho)\leq F_{_2}(s,0)\,$ by the Schwartz inequality,
it follows that the mean scalar energy density is non-increasing.
It is conserved in general only if
$\,\lim\limits_{\rho_0\to0}\,\CP_{_2}(t,\rho_0;s,d\rho)
=\delta(\rho)\,d\rho)$, \,i.e.\,\,if the trajectories are uniquely
determined by their final positions. This is the case for strongly
compressible Kraichnan velocities with $\,\wp>d/\xi^2$. \vskip 1cm

\nsection{5}
\vskip 0.6cm

\noindent ({\small forced scalar advection; intermittent direct
cascade with persistent dissipation and zero mode dominance versus
non-intermittent inverse cascade}) \vskip 0.8cm

\noindent When scalar is forced with the random Gaussian source
characterized by the isotropic 2-point function (\ref{fsr1}) and
if the forcing is independent of velocities and of the initial
distribution of the scalar, then the scalar 2-point function
evolves according to the relation \qq F_{_2}(t,\rho_0)\ =\ \int
F_{_2}(s,\rho)\,\,\CP_{_2}(t,\rho_0;s,d\rho)\ + \
\int\limits_s^td\sigma\int\chi(\rho)\,\,
\CP_{_2}(t,\rho_0;\sigma,d\rho)\,,\label{fod} \qqq which solves
the differential equation \qq {_d\over^{dt}}\,F_{_2}(t,\rho)\ =\
M_2^{\ell=0}\,F_{_2}(t,\rho)\ +\ \chi(\rho)\,. \label{2pe} \qqq
Setting $\,\rho=0\,$ in the latter identity, we obtain the
averaged scalar energy balance with
$\,M_2^{\ell=0}\,F_{_2}(t,\rho)|_{_{\rho=0}}\,$ representing the
mean dissipation rate $\,\epsilon\,$ and $\,\chi(0)\,$ the mean
injection rate of the scalar energy per unit volume. \vskip 0.3cm

The first term in the solution (\ref{fod}) decays with time if the
initial 2-point function $\,F_{_2}(s,\rho)\,$ decays in space. The
second term may be interpreted as the lapse of time between
moments $\,s\,$ and $\,t\,$ when the two trajectories starting at
distance $\,\rho_0\,$ stay in the region of sizable values of the
source covariance $\,\chi$. Its asymptotic behavior depends
crucially on what two Lagrangian trajectories do at long times.
\vskip 0.7cm

\noindent 5.1. \ {\bf Direct versus inverse scalar cascades}
\vskip 0.5cm

\noindent In the weakly compressible regime $\,\wp<d/\xi^2$, \,the
trajectories separate to a fixed distance in a finite time,
Nevertheless, the average time spent by two trajectories within
the range of $\,\chi\,$ is finite only for
$\,\wp<{d-2+\xi\over2\xi}\,$ (or $\,a<1$). In that case, the
scalar 2-point function reaches the stationary form independent of
its initial value \qq \,F_{_2}(\rho)\ =\
\int\limits_{-\infty}^td\sigma\int\chi(\rho)\,\,
\CP_{_2}(t,\rho_0;\sigma,d\rho)\ =\ {_1\over^{D_2(d-1)}}\int
\limits_\rho^\infty\rho_1^{-a}\,d\rho_1\int\limits_0^{\rho_1}
\rho_2^{a-\xi}\,\chi(\rho_2)\,d\rho_2\,, \qqq where the last
equality holds for vanishing $\,\eta$, $\,L\,$ and $\,\kappa$.
\,If $\,{d-2+\xi\over2\xi}\leq\wp<d/\xi^2\,$ then, although the
trajectories separate to a fixed distance in a finite time, with
positive probability they revisit smaller separations. As a
result, the average time they spend within the range of $\,\chi\,$
and, consequently, the scalar 2-point function diverge when
$\,t\to\infty$. \,What still reaches the stationary form, however,
is the 2-point scalar structure function \qq {\bf E}\
[\theta(t,\Nx)-\theta(t,\Nx+\bm\rho)]^2\ \equiv\ S_{_2}(\rho) \ \
\,\mathop{\longrightarrow}\limits_{t\to\infty}\,\ \
{_2\over^{D_2(d-1)}}\int\limits_0^\rho\rho_1^{-a}\,
d\rho_1\int\limits_0^{\rho_1}\rho_2^{a-\xi}\,\chi(\rho_2)\,d\rho_2\,
\qqq which in the limit exhibits the scaling behavior
$\,\simeq\,\rho^{2-\xi}\,$ for small $\,\rho$. \,In the stationary
state, the scalar energy balance reduces to the identity \qq
M_2^{\ell=0}\,F_{_2}(\rho)|_{_{\rho=0}}\ +\ \chi(0)\ =\ 0 \qqq
expressing the fact that the dissipation and injection balance
each other. In particular, the mean dissipation rate $\,\epsilon$,
\,equal for $\,\kappa>0\,$ to $\,2\kappa\,{\bf
E}\,(\bm\nabla\theta)^2$, is $\,\kappa$-independent and \,does not
vanish when $\,\kappa\to0$, the phenomenon called the {\bf
dissipative anomaly}. The anomaly, accompanied by the {\bf direct}
scalar energy {\bf cascade} from the scale on which
$\,\chi(\rho)\,$ decays (where it is injected) to smaller and
smaller scales, is another manifestation of the persistence of the
scalar energy dissipation. It is assured by the explosive
separation of the Lagrangian trajectories in the whole
$\,\wp<d/\xi^2\,$ range. \vskip 0.3cm

In the strongly compressible regime $\,\wp>d/\xi^2$, \,the scalar
2-point function does not stabilize but has a constant
contribution growing linearly in time with the rate equal to
$\,\chi(0)\,$ in the $\,\kappa\to0\,$ limit. No dissipation
persists in this limit due to the deterministic character of the
Lagrangian trajectories. The injected scalar energy ultimately
condenses in the constant mode in the process of the {\bf inverse
cascade} towards longer and longer distances. The 2-point
structure function of the scalar, however, reaches the stationary
form \qq S_{_2}(\rho)\ =\
{_2\over^{D_2(d-1)}}\int\limits_0^\rho\rho_1^{-a}\,
d\rho_1\int\limits_{\rho_1}^\infty\rho_2^{a-\xi}\,
(\chi(0)-\chi(\rho_2))\,d\rho_2\,. \qqq with the scaling behaviors
$\,\propto\rho^{1-a}\,$ for small $\,\rho\,$ and
$\,\propto\rho^{2-\xi}\,$ for large $\,\rho$. \vskip 0.7cm

\noindent 5.2. \ {\bf Zero mode scenario of intermittency}
\vskip 0.5cm

\noindent In the presence of stationary, Gaussian, time
decorrelated sources, the higher-point equal-time correlation
functions $\,{\bf
E}\prod\limits_{n=1}^N\theta(t,\Nx_n)\,\equiv\,F_{_N}(t;\un\Nx)\,$
of scalar satisfy the evolution equations generalizing
(\ref{2pe}): \qq {_d\over^{dt}}\,F_{_{N}}(t,\un\Nx)\ =\
M_{_{N}}\,F_{_{N}}(t,\un\Nx)\ +\
(F_{_{N-2}}\otimes\,\chi)(t,\un\Nx)\,, \label{Npe} \qqq where \qq
(F_{_{N-2}}\otimes\,\chi)(t,\Nx_{_1},\dots,\Nx_{_N})\ =\
\sum\limits_{n<m}F_{_{N-2}}(t,\mathop{\Nx_{_1},.\dots.,\Nx_{_N}}\limits_{
\hat{n}\,\,\,\,\hat{m}})\,\,\chi(\Nx_n-\Nx_m)\,. \qqq The above
relations are solved inductively by the expressions \qq
F_{_N}(t,\un\Nx)\ =\ \int
F_{_N}(s,\un\Ny)\,\,\CP_{_N}(t,\Nx;s,d\Ny)\ + \
\int\limits_s^td\sigma\int
(F_{_{N-2}}\otimes\,\chi)(\sigma,\un\Ny) \
\CP_{_N}(t,\un\Nx;\sigma,d\un\Ny)\,,\label{fodN} \qqq compare to
(\ref{fod}). One expects that for sufficiently weak
compressibility those solutions reach stationary form
$\,F_{_N}(\un\Nx)\,$ vanishing for odd $\,N\,$ and given for even
$\,N\,$ by the second term on the right hand side with
$\,s=-\infty$. \,This has been established rigorously for
$\,\wp=0\,$ and all $\,\kappa\geq0\,$ in (Hakulinen, 2002). \vskip
0.3cm

An important question concerns the behavior at small $\,\rho\,$ of the
$\,\kappa=0\,$ stationary state scalar structure functions
\qq
S_{_{N}}(t,\rho)\ =\ {\bf E}\ [\theta(\Nx)-\theta(\Nx+\bm\rho)]^{N}
\qqq
\noindent with even $\,N$. \,They are given by special combinations
of the correlation
functions $\,F_{_{N}}(\un\Nx)$. \,Naive dimensional predictions based
on eq.\,\,(\ref{Npe})
would suggested the behavior $\,S_{_{2N}}(\rho)\propto\rho^{N(2-\xi)}\,$
since operators $\,M_N\,$ have dimension $\,length^{\xi-2}$. \,This
agrees with the scaling of the 2-point function described above
and would automatically hold for the higher structure functions if the
scalar differences $\,[\theta(t,\Nx)-\theta(t,\Nx+\bm\rho)]\,$ were normally
distributed. The dimensional predictions, first postulated in
(Obukhov, 1949) and (Corrsin, 1951), are in contradiction with
experimental turbulent advection measurements which indicate scaling of
scalar structure functions with exponents that grow slower
than linearly with $\,N$. \,Such behavior of the scaling exponents signals
more frequent appearance than in normal distributions of large fluctuations
of the scalar differences at small separations, the phenomenon called
scalar {\bf intermittency}.
\vskip 0.3cm

It had been suggested in (Kraichnan, 1994) that the higher scalar
structure functions of the Kraichnan model exhibit non-dimensional
scaling. It was subsequently realized in (Shraiman \& Siggia,
1995), (Gaw\c{e}dzki \& Kupiainen, 1995) and (Chertkov, Falkovich,
Kolokolov \& Lebedev, 1995) that, in the incompressible model, the
$\kappa=0\,$ higher point functions $\,F_{_{N}}(\un\Nx)\,$ are
dominated at short distances by the contributions from the scaling
zero modes $\,\varphi_{_N}(\un\Nx)\,$ of the operators
$\,M_{_N}\,$ satisfying \qq M_{_N}\,\varphi_{_N}(\un{\Nx})\ =\
0\,,\qquad\varphi_{_N}
(\lambda\un{\Nx})\,=\,\lambda^{\zeta_{_N}}\,\m\varphi_{_N}
(\un{\Nx})\,. \qqq The scaling dimensions $\,\zeta_{_N}$ of such
modes are not constraint by the dimensional analysis but are
accessible perturbatively or numerically (note that the modes
annihilated by $\,M_{_N}$ drop out in the stationary version of
eq.\,\,(\ref{Npe})). \,The perturbative calculation of such modes
gives for even $\,N$ and general $\,\wp\,$ \qq \zeta_{_N}\ =\
{_N\over^2}(2-\xi)\,-\,{_{N(N-2)(1+2\wp)} \over^{2(d+2)}}\,\xi\ +\
\CO(\xi^2)\,. \qqq In particular, the zero modes dominate at short
distances the structure functions so that \qq S_{_N}(\rho)\
\propto\ \,\rho^{\,\zeta_{_N}}\,. \qqq Such non-dimensional
scaling signals the short distance intermittency of scalar
advected by the weakly compressible Kraichnan model velocities.
\vskip 0.3cm

The {\bf zero mode dominance} of the stationary scalar higher-point
functions has been exhibited by the perturbative
analysis of the Green functions of operators $\,M_{_N}$ around $\,\xi=0$,
$\,d=\infty\,$ and $\,\xi=2$. \,The numerical results gave a similar
picture for all values of $\xi$, see the figure on the next page
representing the values
of the 4-point function anomalous exponent $\,(2\zeta_2-\zeta_4)\,$
obtained by Frisch, Mazzino Noullez \& Vergassola
(1999) in numerical simulations of the three-dimensional (circles)
and 2-dimensional (stars) incompressible Kraichnan model.
\vskip 0.3cm

What is the physical meaning of the zero modes of the operators
$\,M_{_N}$ that dominate the short-distance asymptotics of \m the
scalar $\m N$-point functions? They are {\bf statistically
conserved modes} of the effective diffusion of Lagrangian
trajectories with generators $\,M_{_N}$. \,Indeed, the mean value
of a translationally-invariant scaling function
$\,\psi_{_N}(\un{\Nx})\,$ of scaling dimension $\,\sigma$, viewed
as a function of time $\,t\,$ positions of $\,N\,$ Lagrangian
trajectories, is \qq \int
\psi_{_N}(\un{\Ny})\,\,\CP_{_N}(0,\un{\Nx};t,d\un{\Ny}) \qqq
which, for generic $\,\psi_{_N}$, \,grows dimensionally as
$\,\CO(t^{\sigma/(2-\xi)})\,$ for large $\,t\,$ reflecting the
super-diffusive growth $\,\propto\,t^{1/(2-\xi)}\,$ of the
distances between the trajectories. \,But if
$\,\psi_{_N}=\varphi_{_N}\,$ is a zero mode of $\,M_{_N}$ then the
above expectation is conserved in time (such conserved modes are
accompanied by descendent ones whose Lagrangian averages grow
slower than dimensionally, see (Bernard, Gaw\c{e}dzki \&
Kupiainen, 1997).

\leavevmode\epsffile[40 10 415 474]{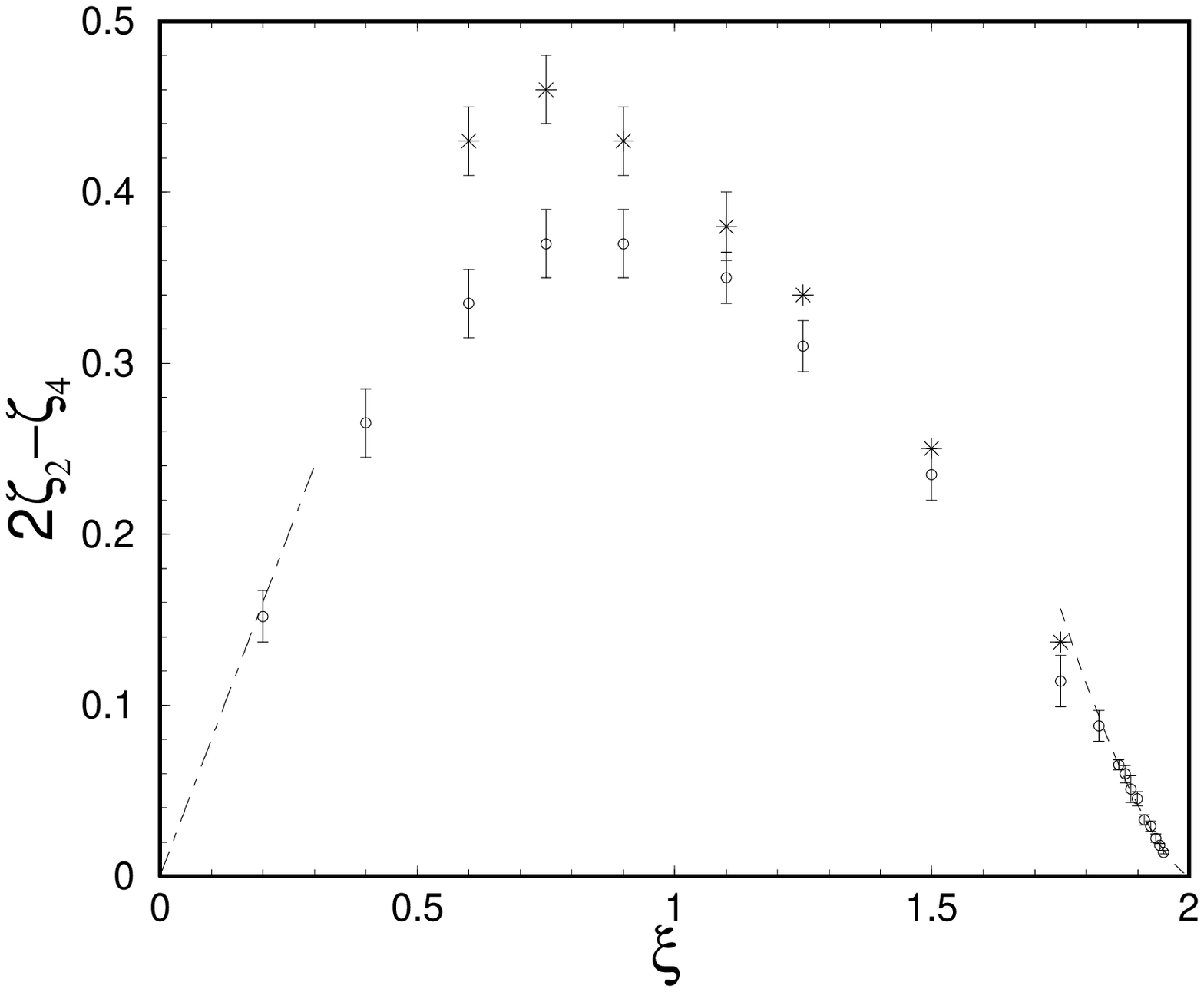}

\noindent 5.3. \ {\bf Non-intermittency of the inverse cascade}
\vskip 0.5cm

\noindent In the strongly compressible phase with the inverse cascade 
of scalar energy, the behavior of the higher structure functions is
different. In fact, only the lower ones stabilize, but the ones
that do, scale normally on large distances. In this regime one can
find exactly the stationary form of the probability density
function of the scalar difference: \qq {\bf E}\
\delta\Big(\vartheta\,-\,{\rho^{-(2-\xi)/2}}\,[\theta(\Nx)\,-\,\theta({\Nx+\bm\rho})]
\Big)\ \ \propto\ \ [\chi(0)\,+\,{\rm const}.
\,\vartheta^2]^{-(a-\xi/2)/(2-\xi)} \qqq at large distances. Its
scaling form indicates that there is no intermittency in the
inverse cascade of the scalar (the deviation from the normal
distribution is scale-independent). For small $\,\rho$, \,however,
all the stabilizing structure functions scale as $\,\rho^{1-a}\,$
signaling an extreme short distance intermittency.

\nconclusions{} \vskip 0.5cm

\noindent As we have seen, the transport of a scalar quantity by 
velocities distributed according to the Kraichnan ensemble shows 
two different phases characterized by different direction of the
scalar energy cascades and different degrees of intermittency. The
phase transition occurs at the value $\wp={d\over\xi^2}$ of the
compressibility degree, where the behavior of the Lagrangian
trajectories changes drastically from the explosive separation to
the implosive aggregation. These two phases are somewhat
reminiscent of the behavior of the three-dimensional versus
two-dimensional developed turbulence. That suggests that one
should put more stress on the Lagrangian methods in studying the
latter, not quite a new lesson, see e.g.\,\,(Pope, 1994), by with
the new twist pointing to the importance of the intrinsically
stochastic character of the Lagrangian flow at extreme Reynolds
numbers. Of course, the Navier-Stokes and the Euler equations,
unlike the scalar advection one, are non-linear, a difference
that, certainly, is far from being minor. Also, they describe
velocity fields that are temporally correlated and transformed
when carried along their own Lagrangian trajectories. Besides, due
to pressure, there are non-local interactions present. Some of
those effects, however, may be studied already in synthetic
velocity ensembles. It seems that the investigation of such
ensembles has a potential to teach us important lessons that have
to be mastered on the way to an understanding of fully developed
turbulence.

\nreferences{} \vskip 0.3cm

\refce Arnold, L. 1998, {\it "Random Dynamical Systems"},
Springer-Verlag, Berlin-Heidelberg-New York

\refce Bernard, D, Gaw\c{e}dzki, K. \& Kupiainen, A. 1998, {\it
``Slow modes in passive advection''}, J. Stat. Phys. {\bf 90},
519-569

\refce Chertkov, M., Falkovich, G., Kolokolov, I. \& Lebedev, V.
1995, {\it ``Normal and anomalous scaling of the fourth-order
correlation function of a randomly advected scalar''}, Phys. Rev.
{\bf E 52}, 4924-4941

\refce Corrsin, S. 1951, {\it ``On the spectrum of isotropic
temperature fluctuations in an isotropic turbulence''}, J. Appl.
Phys. {\bf 22}, 469-473

\refce Falkovich, G., Gaw\c{e}dzki, K., \& Vergassola, M. 2001,
{\it ``Particles and fields in fluid turbulence''}, Rev. Mod.
Phys. {\bf 73}, 913-975

\refce Fannjiang A. \& Papanicolaou, G. C. 1996, {\it ``Diffusion
in turbulence''}, Prob. Theory Rel. Fields, {\bf 105}, 279-334

\refce Fefferman, C. L. 2000, {\it ``Existence \& smoothness of
the Navier-Stokes equation''}, millennium prize problem
description, http://www.claymath.org/prizeproblems/navier$\un{\
}$stokes.pdf

\refce Frisch, U., Mazzino, A., Noullez, A. \& Vergassola, M.
1999, {\it ``Lagrangian method for multiple correlations in
passive scalar advection''}, Phys. Fluids {\bf 11}, 2178-2186

\refce Gaw\c{e}dzki, K. \& Kupiainen, A. 1995, {\it ``Anomalous
scaling of the passive scalar}, Phys. Rev. Lett. {\bf 75},
3834-3837

\refce Gaw\c{e}dzki, K. \& Vergassola, M. 2000, {\it ``Phase
transition in the passive scalar advection''}, Physica D {\bf
138}, 63-90

\refce Hakulinen, V. 2002, {\it ``Passive advection and degenerate
elliptic operators $M_n$''}, PhD thesis, http://
ethesis.helsinki.fi/julkaisut/mat/matem/vk/hakulinen/passivea.pdf

\refce Kolmogorov, A. N. 1941, {\it ``The local structure of
turbulence in incompressible viscous fluid for very large
Reynolds' numbers''}, C. R. Acad. Sci. URSS {\bf 30}, 301-305

\refce Kraichnan, R. H. 1968, {\it ``Small-scale structure of a
scalar field convected by turbulence''}, Phys. Fluids {\bf 11},
945-963

\refce Kraichnan, R. H. 1994, {\it ``Anomalous scaling of a
randomly advected passive scalar''}, Phys. Rev. Lett. {\bf 72},
1016-1019

\refce Le Jan, Y. \& Raimond, O. 1998, {\it ``Solution
statistiques fortes des \'{e}quations diff\'{e}rentielles
stochastiques''}, arXiv:math.PR/9909147

\refce Majda, A. J. \& Kramer, P. R. 1999, {\it ``Simplified
models for turbulent diffusion: theory, numerical modeling, and
physical phenomena''}, Phys. Rep. {\bf 314}, 237-574

\refce Obukhov, A. M. 1949, {\it ``Structure of the temperature
field in a turbulent flow''}, Izv. Akad. Nauk SSSR, Geogr. Geofiz.
{\bf 13}, 58-69

\refce Pope, S. B. 1994, {\it ``Lagrangian pdf methods for
turbulent flows''}, Ann. Rev. Fluid Mech. {\bf 26}, 23-63

\refce Richardson, L. F. 1926, {\it ``Atmospheric diffusion shown
on a distance-neighbour graph''}, Proc. R. Soc. Lond. {\bf A 110},
709-737

\refce Shraiman, B. I. \& Siggia, E. D. 1995, {\it ``Anomalous
scaling of a passive scalar in turbulent flow''}, C.R. Acad. Sci.
{\bf 321}, 279-284

\refce Shraiman, B. I. \& Siggia, E. D. 1996, {\it ``Symmetry and
scaling of turbulent mixing''}, Phys. Rev. Lett. {\bf 77},
2463-2466

\refce Taylor, G. I. 1921, {\it ``Diffusion by continuous
movements''}, Proc. London Mat. Soc. {\bf 20}, 196-212

\end{document}